\documentclass[twocolumn]{aastex7}

\newcommand{\ctext}[1]{\raise0.2ex\hbox{\textcircled{\scriptsize{#1}}}}
\newcommand{\MBH}{$M_{\rm{BH}}$ }

\newcommand{\Mstar}{$M_{\star}$ }

\usepackage{comment}
\usepackage{afterpage}
\usepackage{here}
\usepackage{threeparttable}
\usepackage[]{lineno}
\usepackage{booktabs}

\usepackage{tikz}
\usetikzlibrary{shapes,arrows,positioning}

\submitjournal{ApJ}
\shorttitle{Environment of quasars at $z=0.4$--$1.0$}
\shortauthors{Shibata et al.}

\begin{document}

\title{Environment of SDSS quasars at $z=0.4$--$1.0$ explored  by Subaru HSC}

\author[0009-0001-8283-6308]{Kohei Shibata}
\affiliation{Graduate School of Science and Engineering, Ehime University, 2-5 Bunkyo-cho, Matsuyama, Ehime 790-8577, Japan} 
\email[show]{shibata@cosmos.phys.sci.ehime-u.ac.jp} 

\author{Tohru Nagao}
\affiliation{Research Center for Space and Cosmic Evolution, Ehime University, 2-5 Bunkyo-cho, Matsuyama, Ehime 790-8577, Japan}
\affiliation{Amanogawa Galaxy Astronomy Research Center, Kagoshima University, 1-21-35 Korimoto, Kagoshima 890-0065, Japan}
\email{tohru@cosmos.phys.sci.ehime-u.ac.jp}

\author[0000-0002-0673-0632]{Hisakazu Uchiyama}
\affiliation{National Astronomical Observatory of Japan, 2-21-1 Osawa, Mitaka, Tokyo 181-8588, Japan}
\affiliation{Department of Advanced Sciences, Faculty of Science and
Engineering, Hosei University, 3-7-2 Kajino-cho, Koganei, Tokyo
184-8584, Japan}
\email{hisakazu.uchiyama.86@hosei.ac.jp}

\author[0000-0002-7598-5292]{Mariko Kubo}
\affiliation{School of Science, Kwansei Gakuin University, Sanda, Hyogo 669-1337, Japan}
\affiliation{Astronomical Institute, Tohoku University, 6-3, Aramaki, Aoba-ku, Sendai, Miyagi, 980-8578, Japan}
\email{markubo@kwansei.ac.jp}

\author[0000-0002-3531-7863]{Yoshiki Toba}
\affiliation{Department of Physical Sciences, Ritsumeikan University, 1-1-1 
Noji-higashi, Kusatsu, Shiga 525-8577, Japan}
\affiliation{National Astronomical Observatory of Japan, 2-21-1 Osawa, Mitaka, Tokyo 181-8588, Japan}
\affiliation{Academia Sinica Institute of Astronomy and Astrophysics, 11F of Astronomy-Mathematics Building, AS/NTU, No.1, Section 4, Roosevelt Road, Taipei 10617, Taiwan}
\affiliation{Research Center for Space and Cosmic Evolution, Ehime University, 2-5 Bunkyo-cho, Matsuyama, Ehime 790-8577, Japan}
\email{toba@fc.ritsumei.ac.jp}

\author[0000-0002-8432-6870]{Kiyoaki Christopher Omori}
\affiliation{Research Center for Space and Cosmic Evolution, Ehime University, 2-5 Bunkyo-cho, Matsuyama, Ehime 790-8577, Japan}
\affiliation{Department of Astronomy and Physics and Institute for Computational Astrophysics, Saint Mary's University, 923 Robie St, Halifax, NS B3H 3C3, Canada}
\email{k.omori116@gmail.com}

\author{Toshihiro Kawaguchi}
\affiliation{Department of Economics, Management and Information Science, Onomichi City University, Hisayamada 1600-2, Onomichi, Hiroshima 722-8506, Japan}
\email{kawaguchi@onomichi-u.ac.jp}

\author[0009-0000-2525-9236]{Yuta Suzuki}
\affiliation{Graduate School of Science and Engineering, Ehime University, 2-5 Bunkyo-cho, Matsuyama, Ehime 790-8577, Japan}
\email{yuta@cosmos.phys.sci.ehime-u.ac.jp}


\begin{abstract} 
The relationship between quasars and their galaxy environment is important for understanding the evolution of galaxies and supermassive black holes, but it is not fully understood. 
We perform a wide and deep exploration of the environment of quasars at $0.4 < z < 1.0$ using the Hyper Suprime-Cam Subaru Strategic Program (HSC-SSP) survey. 
We investigate the environment of the 1,912 spectroscopically selected quasars from the Sloan Digital Sky Survey (SDSS), using photometrically selected galaxies from the HSC-SSP data, over an area of 505 deg${^{2}}$.
The quasar environment is compared to the environment of matched galaxies with similar stellar mass and redshift. 
We employ the $k$-nearest neighbor method to define the local galaxy number density for both the quasars and the matched galaxies at a scale of a few hundred kpc. 
As a result, we find that the number density of galaxies around SDSS quasars is lower than that of the matched galaxies by $\sim$11--$20\%$.
We also investigate possible correlations between the local galaxy number densities and the quasar properties such as black hole mass and Eddington ratio.
As a result, no correlation is found between the local galaxy number densities and these properties of quasars.
These results suggest that the quasar activity is not triggered by the high number density of surrounding galaxies at the scale of a few hundred kpc.

\end{abstract}



\section{Introduction} \label{sec:intro}
Understanding the evolution of galaxies is one of the major themes in modern astronomy.
Research over the past few decades has revealed that massive galaxies ubiquitously host a supermassive black hole (SMBH) at their center \citep[e.g.,][]{Ghez08, Genzel10, Event19}.
Observations of massive galaxies have shown a strong correlation between the mass of central SMBHs and the properties of the spheroids, suggesting that SMBHs and their host galaxies have co-evolved \citep[e.g.,][]{Magorrian98, Marconi03, Kormendy13}.
Therefore, understanding the mass growth of SMBHs is essential for fully understanding the galaxy evolution.

Quasars are among the most luminous objects in the Universe, and their luminosity is powered by mass accretion onto SMBHs at the center of their host galaxies \citep{Salpeter64, Lynden69}.  
To deliver gas from host galaxies to the vicinity of SMBHs, the gas needs to lose most of its angular momentum \citep[e.g.,][]{Jogee06}, which is challenging but necessary process to trigger the quasar activity.  
One proposed mechanism to explain the efficient gas accretion onto SMBHs is the ``major merger scenario'' \citep[e.g.,][]{Sanders88, Hopkins08}, where the merging of gas-rich galaxies leads to the angular momentum transfer, resulting in the gas accretion onto the SMBH.

The frequency of galaxy mergers is known to be related to the surrounding environment. 
For instance,  some studies reported that galaxy mergers are more likely to occur in the high-density environment \citep[e.g.,][]{de11, Pearson24}, while other studies suggested that mergers are hard to occur frequently in the high-density environment because the high velocity dispersion prevents galaxy mergers \citep[e.g.,][]{Omori23, Sureshkumar24}. 
Although there is no consensus on the types of environment that favor galaxy mergers, these findings indicate a connection between galaxy mergers and the surrounding environment.

To test the major merger scenario observationally, various studies have examined the environment around quasars.
In \cite{Serber06}, the environment of 2,028 quasars selected from the Sloan Digital Sky Survey (SDSS; \citealt{York00, Schneider05}) at $0.08 \leq z \leq 0.4$ was examined from 25 kpc to 1 Mpc around the quasars. 
Based on their study, quasars are located in higher-density environment compared to the environment of $L^\ast$ galaxies.
Specifically, the quasar environment was found to be denser within a radius of 100 kpc by a factor of $\sim$ 1.4 compared to that of typical $L^\ast$ galaxies.
In \cite{Falder10}, the surrounding environment of 173 active galactic nuclei (AGNs; but dominated by quasars) at $z \sim 1$ was investigated using infrared data from the Spitzer Space Telescope. 
This study concluded that the environment of AGNs within 1 Mpc is significantly denser compared to that of blank fields. 
Especially, within 300 kpc, the AGN environment exhibited an especially high density, reaching 8.5 $\sigma$ significance, with the number density of galaxies gradually decreasing until reaching Mpc scales.
\cite{Zhang13} utilized data from the Stripe 82 region of the SDSS to investigate the environment of 2,300 quasars at redshifts between $0.6 < z < 1.2$. They reported that quasars show an overdensity of galaxies compared to the background field galaxies and that the clustering amplitude increases as redshift increases. 
Such overdensity properties around quasars have been reported also at high-$z$ ($z>2$) \citep[e.g.,][]{Kashikawa07, Utsumi10, Falder11,  Pudoka24}.


On the other hand, some observational studies have reported different trends. 
\cite{Coldwell06} compared the environment of 1,652 SDSS quasars at $0.1 < z < 0.2$ with that of 1,153 non-active galaxies and 102 galaxy groups. They found that quasars and non-active galaxies live in a similar environment, which is significantly less dense than that of galaxy groups.
\cite{Karhunen14} examined the environment of 302 SDSS quasars in the Stripe 82 region with $z < 0.5$. They investigated the environment at scales of 200 kpc to 1 Mpc and reported no significant difference between quasars and comparison galaxies.
\cite{Bettoni15} conducted an environmental study on 52 quasars from \cite{Falomo14} with detection of host galaxies in the SDSS Stripe 82 region at $z < 0.3$, examining the surrounding environment within 50 kpc. They created a control sample matched in terms of the absolute magnitude of quasar host galaxies and redshift distribution. By comparing the number count of companion galaxies within 50 kpc, they found no significant difference between the quasar sample and the control sample.
\cite{Wethers22} studied the environment of 205 quasars at $0.1 < z < 0.35$ using data from the Galaxy and Mass Assembly (GAMA) survey. They compared the environment within $< 100$ kpc, sub-Mpc, and several Mpc scales with those of comparison galaxies and found no significant differences at any scales.
Such a lack of overdensity properties around quasars also in the high-$z$ ($z>2$) Universe has been suggested by some observational studies \citep[e.g.,][]{Kashikawa07, Goto17, Kikuta17, Uchiyama18, Lambert24, Suzuki24}.


As shown above, the environment around quasars is still controversial. 
These factors, including the rarity of quasars, variations in the definition of surrounding galaxy density, and differences in observational depth and field of view, contribute to the lack of statistical consensus.
Therefore, in this study, we aim to resolve the debate on whether the quasar lives in overdense regions, by conducting a detailed statistical investigation of the surrounding galaxy environment of quasars, with a focus on low-$z$ quasars, which allow for a more detailed investigation of their environment than high-$z$ quasars.
For this purpose, we utilize the wide and deep broad-band photometric data of galaxies obtained from the Hyper Suprime-Cam (HSC; \citealt{Miyazaki12, Miyazaki18, Furusawa18, Kawanomoto18, Komiyama18}), a prime-focus camera with a field-of-view of 1.5 degree diameter on the Subaru Telescope.

The paper is organized as follows. In Section \ref{Data}, we describe the data of HSC and SDSS. 
In Section \ref{Measurement of Galaxy Environment}, the method to measure the local galaxy number density around quasars and control galaxies is explained.
In Section \ref{Results}, we show the results of the local galaxy number density around quasars and control galaxies.
In Section \ref{Discussion},  we discuss the implications of our results. Specifically, we show the correlations between the local galaxy number densities around quasars and their properties, and discuss the origin of these correlations.
Finally, in Section \ref{Conclusion}, we summarize our findings.
We adopt the following cosmological parameters; 
$\Omega_{M} = 0.27 $, $\Omega_{\Lambda} = 0.73$, and $H_{0} = 70 ~ $km s$^{-1}$ Mpc$^{-1}$.  Magnitudes are given in the AB system \citep{Oke83}. 

\section{Data}\label{Data}
\subsection{HSC galaxy sample for the density measurement}\label{sec:2.1} 
The HSC Subaru Strategic Program (HSC-SSP; \citealt{Aihara18}) is a wide-field optical imaging survey using HSC.
The HSC-SSP consists of three layers: Wide, Deep, and Ultra Deep.
In this study, we use the Wide layer data of the internal data release (DR) S20A \citep{Aihara22}, which consists of $\sim1,100$ deg${^{2}}$ wide-field images with median seeing of $0\, \farcs6 - 0\, \farcs8$ and was taken by the five optical filters of $grizy$ bands \citep{Kawanomoto18}.
The 5$\,\sigma$ limiting magnitudes of the $grizy$ bands are 26.5, 26.5, 26.2, 25.2, and 24.4, respectively \citep{Aihara22}.

The HSC-SSP provides multiple photometric-redshift (photo-$z$) catalogs \citep{Tanaka18}. In this study, we use an HSC photo-$z$ catalog constructed by using spectral energy distribution (SED) fitting code with Bayesian physical priors, Mizuki \citep{Tanaka15, Tanaka18}. See \cite{Tanaka15} and \cite{Tanaka18} for more details. 

In order to measure the local galaxy number density around quasars, we apply some flags to obtain a clean photometric sample (for details, see Appendix \ref{Ap:A}). 
We select 15,465,043 galaxies with $z=0.2$--$1.2$ and $\texttt{i\_cmodel\_mag} - \texttt{a\_i} \leq 23.5$ from the clean photometry sample, where  $\mathrm{i\_cmodel\_mag}$ and \texttt{a\_i} are composite model  $i$-band magnitude and galactic extinction at $i$-band (hereinafter referred to as ``parent galaxy sample''). 
The mean and standard deviation of redshift, stellar mass, and $i$-band absolute magnitude of the parent galaxy sample are $z = 0.73\pm0.24$, $\mathrm{log(}M_{\star}/M_{\odot}) = 10.0\pm0.63$, and $M_{i} = -21.2\pm1.20\  \rm{mag}$, respectively.
The redshift, $M_{\star}$, and $M_{i}$ we use are estimated by Mizuki.

\begin{figure}
 \begin{center}
  \includegraphics[width=1.0\linewidth]{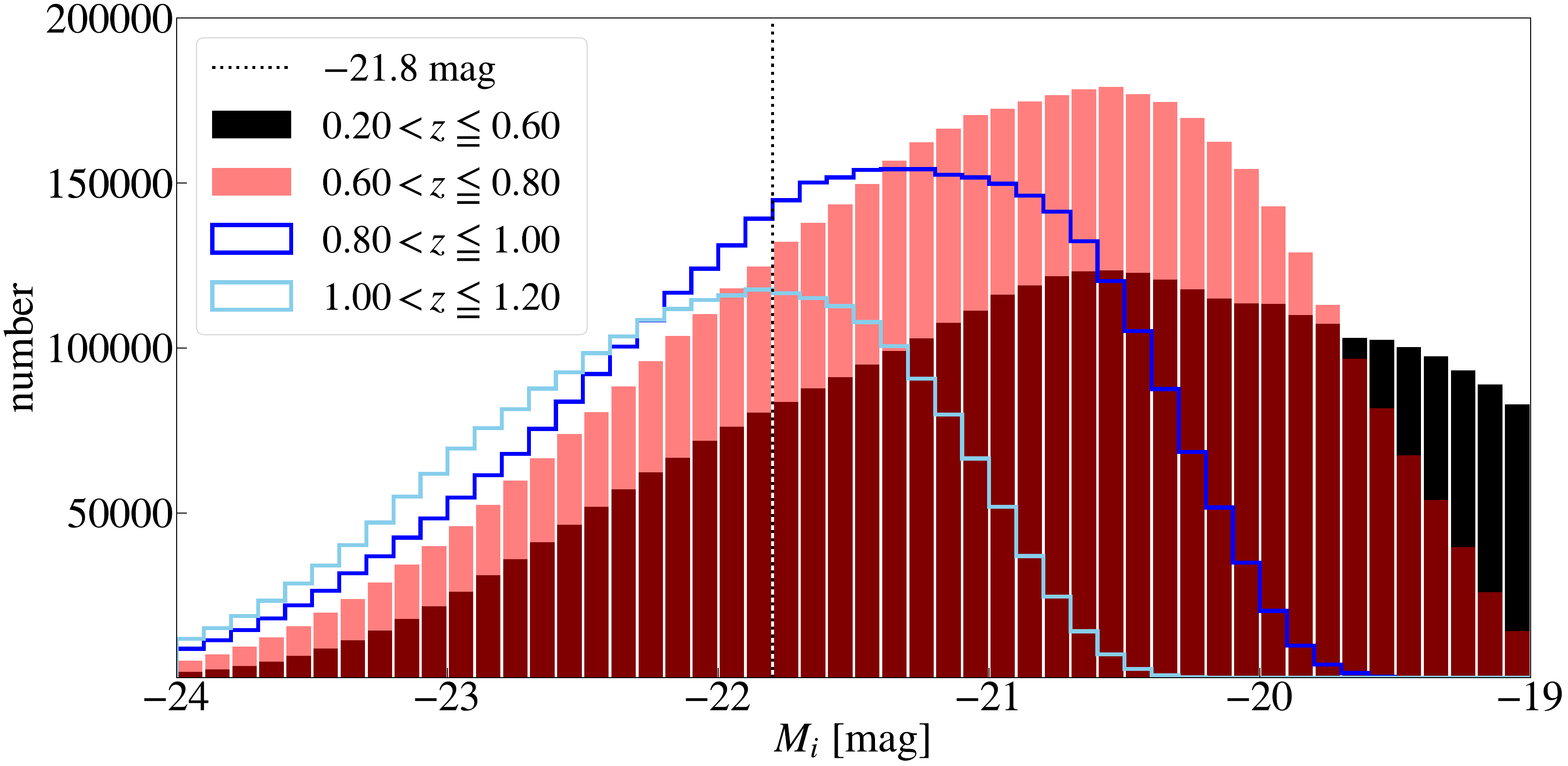}
  \caption{Distributions of the $i$-band absolute magnitudes of the parent galaxy sample with each redshifts. Black, pink, blue and skyblue histograms denote the galaxy number in the redshift bin of the $0.2 < z \leq 0.6$, $0.6 < z \leq 0.8$, $0.8 < z \leq 1.0$, and $1.0 < z \leq 1.2$, respectively. The dotted line indicates $-21.8$ mag.}
  \label{fig:maglimit}
 \end{center}
\end{figure}

\begin{figure*}
 \begin{center}
  \includegraphics[width=0.7\linewidth]{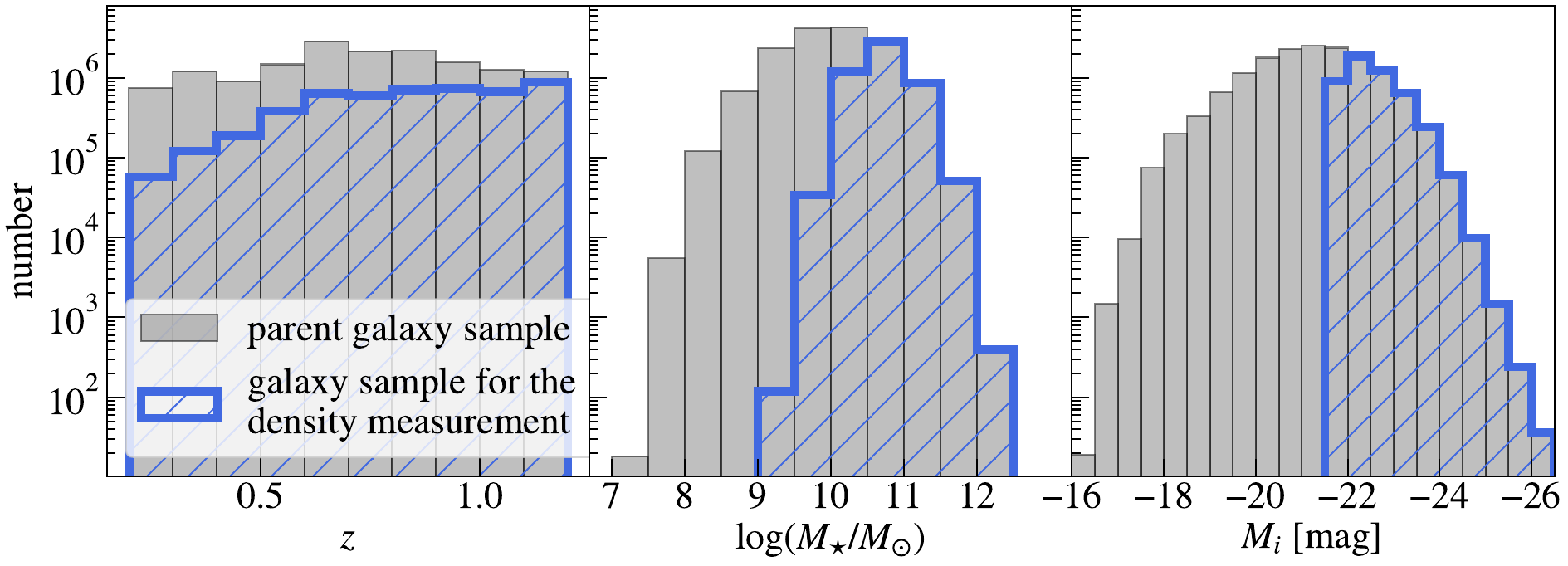}
  \caption{Distributions of redshift (left), stellar mass (center), and $i$-band absolute magnitude (right) for the parent galaxy sample (gray) and the galaxy sample for density measurement (blue).}
  \label{fig:gal_distribution}
 \end{center}
\end{figure*}

In order to remove the redshift bias in the measurement of the quasar environment, we restrict the $i$-band absolute magnitude of the parent galaxy sample.
Figure~\ref{fig:maglimit} shows the $i$-band absolute magnitude distribution of galaxies in the parent galaxy sample, divided into 4 redshift ranges. This figure suggests that the selection of galaxies is highly incomplete for $M_{i} > -21.8$ in the highest redshift range, $1.0 < z \leq 1.2$.
Therefore we construct a galaxy sample for the density measurement by taking galaxies with $M_{i} \leq -21.8$ from the parent galaxy sample (hereinafter ``galaxy sample for the density measurement''). 
Consequently, there are 4,958,316 galaxies in the galaxy sample for the density measurement.
The mean and standard deviation of redshift, stellar mass, and $i$-band absolute magnitude of the galaxy sample for the density measurement are $z = 0.85\pm0.23$, $\mathrm{log(}M_{\star}/M_{\odot}) = 10.7\pm0.32$, and $M_{i} = -22.5\pm0.55$, respectively.
Figure~\ref{fig:gal_distribution} shows distributions of redshift, stellar mass, and absolute $i$-band magnitude of the galaxy sample for the density measurement and the parent galaxy sample.
This figure indicates that the galaxy sample for density measurement is brighter and more massive than the parent galaxy sample.

\subsection{HSC random point}\label{sec:random}
To accurately measure the environment, it is necessary to account for regions that are masked due to bright stars or other contaminants. For this purpose, we utilize random points from the HSC random point catalog \citep{Coupon18} in the DR S20A. These random points are selected with the same flag criteria applied to the HSC photometric galaxies, ensuring consistency in the selection process.

The random points are distributed with a density of 100 points per square arcmin \citep{Coupon18}, providing a robust basis for estimating the masked regions. 
By utilizing this random point catalog with some adequate flags, we estimate $C_{k}$, the fraction of the unmasked area within the radius $d_k$, defined as follows:
\begin{equation} 
\label{eq:def_C}
C_{k} = \frac{N}{\pi {d_{k}}^2\, [\rm{arcmin}^2] \times 100\, [\rm{arcmin}^{-2}]}, 
\end{equation} 
where $N$ represents the number of random points within a radius $d_{k}$. This approach allows us to systematically account for the effects of unmasked regions and ensures the accuracy of our environmental measurements.

Furthermore, by counting the number of the selected random points, we calculate the total survey area for this study. As a result, we determine the effective area of this study to be $\sim 505~\mathrm{deg}^2$. 

\subsection{SDSS quasar sample}\label{sec:2.2}
\begin{figure*}
 \begin{center}
   \includegraphics[width=0.9\linewidth]{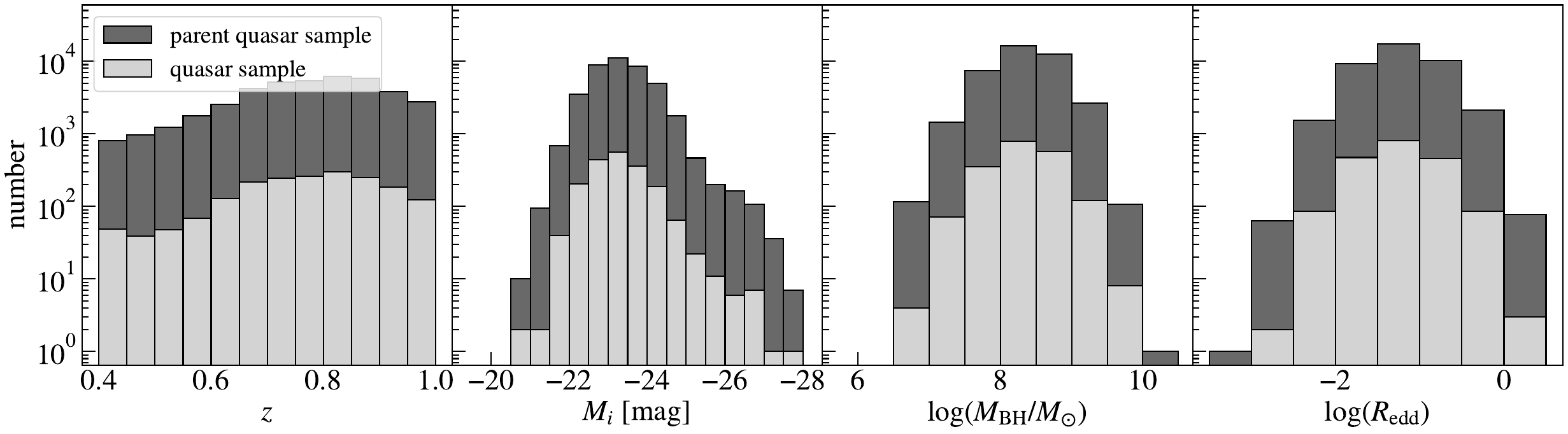}
  \caption{Distributions of the redshift (leftmost panel), $i$-band absolute magnitude (left-center panel), SMBH mass (right-center panel), and Eddington ratio (rightmost panel), for the parent quasar sample (dark gray, 40,664 objects) and quasar sample (light gray, 1,912 objects).}
  \label{fig:qso_distribution}
 \end{center}
\end{figure*}

The quasar sample used in this study is taken from the SDSS DR14 quasar catalog \citep{Paris18}, which contains 526,357 spectroscopically confirmed quasars.
The physical properties of the SDSS quasars used in this study are taken from \cite{Rakshit20}.
In order to statistically investigate the quasar environment, we extend the redshift range of quasars up to $z=$1.0, which has not been examined much in previous studies.
We extract a clean sample from \cite{Rakshit20} with $z=0.4$--$1.0$ (for details, see Appendix \ref{Ap:B}), which consists of 40,664 quasars (hereinafter ``parent quasar sample''), selecting quasars for which the SMBH mass is accurately estimated from a broad line. 
From this parent quasar sample, we select 4,429 quasars from the region covering the HSC survey area.
Furthermore, we exclude 540 quasars whose selection criteria for the SDSS spectroscopy are different from others (see Appendix \ref{Ap:B}), since quasars with largely different selection criteria may affect the statistics of the environment. 
Based on these criteria, we select 3,889 quasars for this study.
For these 3,889 quasars, we measure the local galaxy number densities $\Sigma_{k}$ with $k=2, 5$, and 10 (see Section \ref{Measurement of Galaxy Environment}). 
Additionally, we exclude the quasars with $C_{k} < 0.5$ at all $k\ (=2, 5, 10)$.
There are 1,912 quasars at this point (hereinafter referred to as ``quasar sample'').
Figure~\ref{fig:qso_distribution} shows the distribution of the redshift, $i$-band absolute magnitude ($M_i$), SMBH mass ($M_{\mathrm{BH}}$), and Eddington ratio ($R_{\mathrm{Edd}}$), for the quasar sample.
Table~\ref{tbl:parentqso} shows the mean and median of the redshift, $i$-band absolute magnitude, SMBH mass, and Eddington ratio, for the quasar sample and the parent quasar sample. Figure~\ref{fig:qso_distribution} and Table~\ref{tbl:parentqso} indicate that the quasar sample is not significantly different from the parent quasar sample.

\begin{table}
  \begin{center}
    \caption{Mean and median of the redshift, $i$-band absolute magnitude, SMBH mass, and Eddington ratio for the parent quasar sample and the quasar sample.}
    \label{tbl:parentqso}
    \begin{tabular}{lcccc} 
      \hline \hline
      & \multicolumn{2}{c}{parent quasar sample}                       &\multicolumn{2}{c}{quasar sample} \\ \cline{2-5}
										& mean		& median 	& mean 		&median 	\\ \hline
      ${z}$									&$0.77$		&$0.78$		&$0.77$		&$0.78$	\\
      $M_{i}$								&$-23.38$	&$-23.32$	&$-23.30$	&$-23.24$	\\
      log(${M_{\mathrm{BH}}/M_{\odot}}$)		&$8.34$		&$8.36$		&$8.33$		&$8.34$	\\ 
      log($R_{\mathrm{Edd}}$)				&$-1.22$	&$-1.23$	&$-1.25$	&$-1.26$\\ \hline
    \end{tabular}
  \end{center}
\end{table}

\subsection{matched galaxy sample}\label{sec:2.3}
To directly compare the environments of quasars to non-active galaxies, we construct a one-to-one matched sample of galaxies with similar stellar mass and redshift using the following procedure.
\begin{enumerate}
  \item The stellar mass of quasar host $M_{\star- \rm{qso}}$ is estimated from the SMBH mass \MBH\ \citep{Marconi03, Haring04},
  \begin{equation}\label{eq:MBH}
M_{\star- \rm{qso}} = \frac{M_{\rm{BH}}}{1.75 \times 10^{-3}} .
\end{equation}

  \item In order to  select the matched-redshift and matched-\Mstar galaxy, a quantity $\Delta C$ using the \Mstar and \MBH \ \citep{Wethers22} is defined as follows,
   \begin{equation}
\Delta C = \left(\frac{z-z_{\rm{qso}}}{0.01}\right)^{2} + \left(\frac{\rm{log(}\it{M}_{\star})-\rm{log(}\it{M}_{\star-\rm{qso}})}{0.1}\right)^{2},
\end	{equation}
where $z_{\rm{qso}}$, $M_{\star-\rm{qso}}$, $z$, $M_{\star}$ are redshift of a quasar, stellar mass of a quasar host, redshift of a galaxy, and stellar mass of a galaxy, respectively.
Then, we select the galaxy with smallest $\Delta C$ as a matched galaxy candidate for each quasar.

  \item We measure the local galaxy number density of the matched galaxy candidate. In this time, if the $C_{k} < 0.5$ at all $k\ (=2, 5, 10)$ of the matched galaxy candidate (see Section \ref{Measurement of Galaxy Environment} for $k$ and Section \ref{sec:random} for $C_k$), we exclude this object and reselect another matched galaxy candidate.
\end{enumerate}
Finally, we select the 1,912 matched galaxies (hereinafter referred to as ``matched galaxy sample'').

\begin{table}
  \begin{center}
    \caption{Mean and standard deviation of the redshift, $i$-band absolute magnitude, and stellar mass, for the quasar sample and the matched galaxy sample.}
    \label{tbl:qso-matched}
    \begin{tabular}{lcccc} 
      \hline \hline
      & \multicolumn{2}{c}{quasar sample}                       &\multicolumn{2}{c}{matched galaxy sample} \\ \cline{2-5}
										& mean		& std 		& mean 		&std 	\\ \hline
      ${z}$									&$0.77$		&$0.13$		&$0.77$		&$0.13$	\\
      $M_{i}$								&$-23.30$	&$0.77$		&$-23.17$	&$0.98$	\\
      log(${M_{\star}/M_{\odot}}$)			&$11.09$		&$0.46$		&$11.09$		&$0.45$	\\ \hline
    \end{tabular}
  \end{center}
\end{table}

\begin{figure*}
 \begin{center}
  \includegraphics[width=0.7\linewidth]{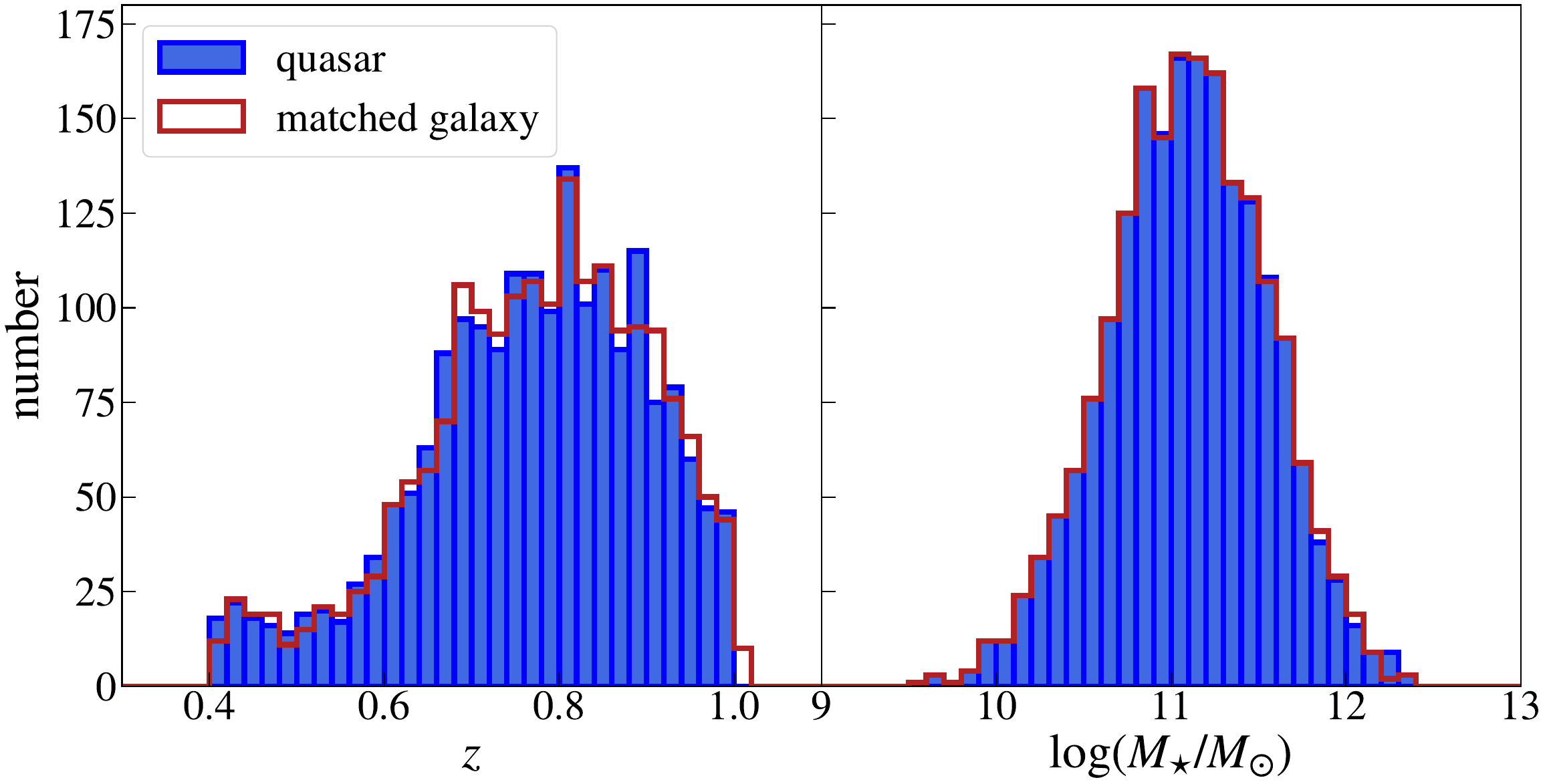}
  \caption{Distributions of the redshift (left panel) and the stellar mass (right panel) for quasar sample (blue bar) and matched galaxy sample (thick red line).}
  \label{fig:hist_qso_matched}
 \end{center}
\end{figure*}

Figure~\ref{fig:hist_qso_matched} shows the distribution of redshift and stellar mass  for the quasar sample and the matched galaxy sample.
Table~\ref{tbl:qso-matched} shows the value of mean and standard deviation for redshift, absolute $i$-band magnitude, and stellar mass of the quasar sample and the matched galaxy sample. Figure~\ref{fig:hist_qso_matched} and Table~\ref{tbl:qso-matched} show no significant difference in the distributions between the quasar sample and the matched galaxy sample. To check the possible systematic difference in the distributions more quantitatively, we apply the Kolmogorov-Smirnov (KS) statistical test. As a result, the $p$-values for the distributions of the redshift and the stellar mass between quasar sample and matched galaxy sample are $0.77$ and $1.00$, respectively, indicating no statistically significant differences in the redshift and stellar-mass distributions between the two samples. 
In Table~\ref{tbl:sample}, we summarize the samples use in this study.

\begin{table}
  \begin{center}
    \caption{Number of objects in each of samples.}
    \label{tbl:sample}
    \begin{tabular}{lr} 
      \hline \hline
      \text{Name of sample}                     & \text{$N$}  \\ \hline
      parent quasar sample            					& 40,664            \\
        quasar sample            							& 1,912           \\
        parent galaxy sample            					& 15,465,043     \\   
        galaxy sample for density measurement            	& 4,958,316        \\
        matched galaxy sample            					& 1,912          \\ \hline
    \end{tabular}
  \end{center}
\end{table}

\section{Measurement of Galaxy Environment} \label{Measurement of Galaxy Environment}
We adopt the $k$-nearest neighbor method \citep{Cooper05, Muldrew12} to measure the local galaxy number densities as local galaxy environments of the quasars and matched galaxies.
The details are as follows.
The local galaxy number density of a target object (a quasar or matched galaxy), $\Sigma _{k}$ is calculated by the following equation:
\begin{equation} 
\Sigma_{k}\equiv \frac{k}{C_{k} \pi d_{k}^2} ~~ \mathrm{pMpc}^{-2}, \label{eq:3}
\end{equation}
where $d_{k}$ [pMpc] is the projected distance from a target to the $k$-th nearest neighbor and $C_{k}$ is the fraction of the unmasked area within the radius $d_{k}$ (\citealt{Uchiyama22}, see Section \ref{sec:random} for $C_k$).
So, $C_{k} \pi d_{k}^2$ means the effective area within the radius $d_{k}$.
Note that, for the targets whose surrounding region is mostly masked, we cannot correctly measure the environment and therefore we do not use it in the statistical analysis in this work.
Specifically, for each value of $k$, we require  $C_{k} \geq 0.5$.
Quasars that do not satisfy $C_{k} \geq 0.5$ for all $k$ are excluded from the quasar sample.

\begin{figure}
 \begin{center}
  \includegraphics[width=0.85\linewidth]{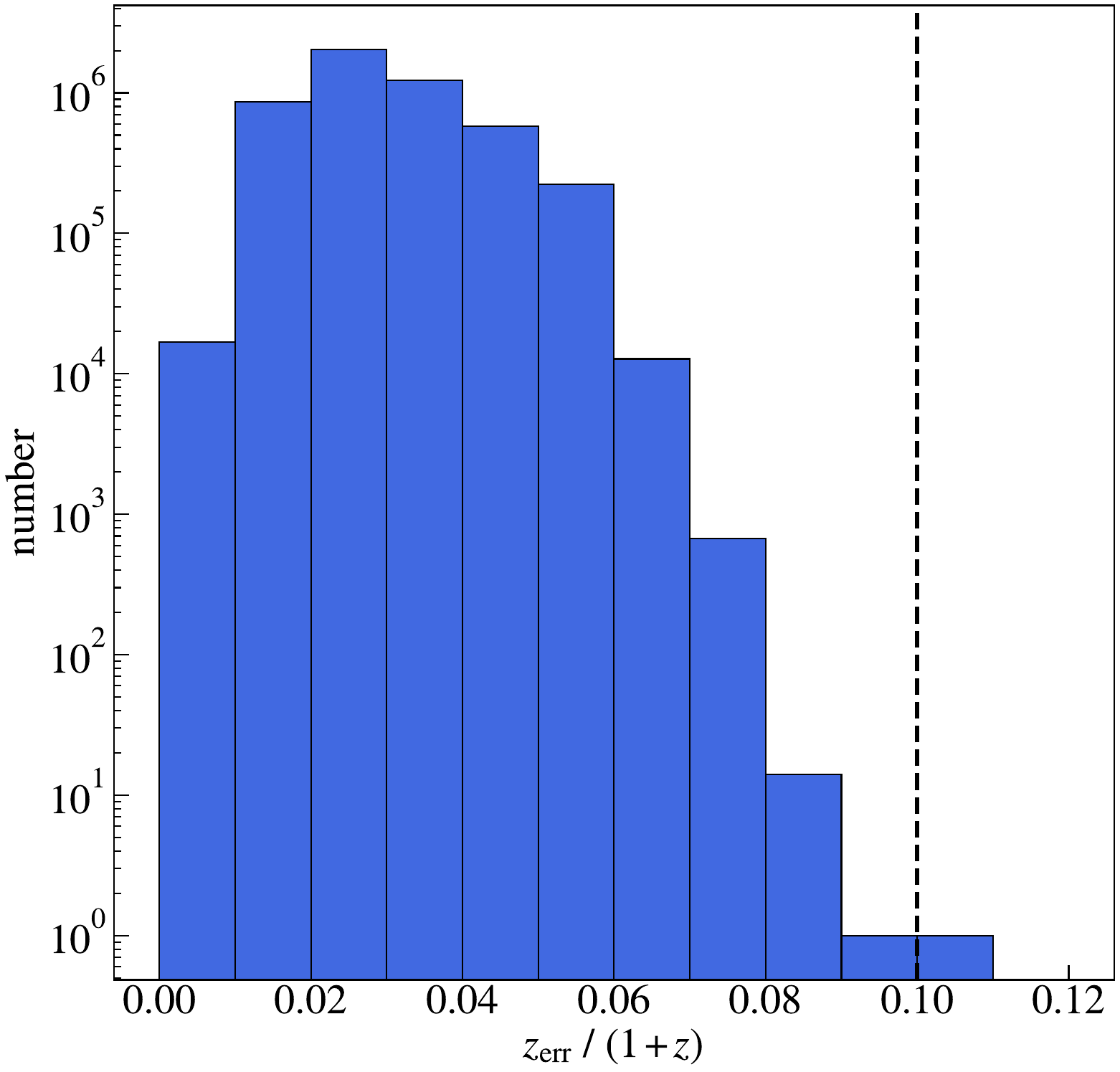}
  \caption{Histogram of the ratio of the photometric error and ($1 + z$) of the galaxy sample for the density measurement. The dashed line denotes to $z_{\mathrm{err}}/(1+z) = 0.1$.}
  \label{fig:photo-z_err}
 \end{center}
\end{figure}

In order to accurately measure the surface galaxy number density without serious effects by foreground or background galaxies, we impose the following redshift restrictions \citep{Kolwa19};
\begin{equation} 
\label{eq:z_slice}
z_{\mathrm{target}} - \Delta z  < z_{\mathrm{galaxy}} < z_{\mathrm{target}}  + \Delta z,
\end{equation}
\begin{equation}
  \Delta z \equiv 0.1 \times (1+z_{\mathrm{target}}).
\end{equation}
If the size of the redshift slice in Equation (\ref{eq:z_slice}) is smaller than a galaxy's photometric redshift error, some galaxies around the target at similar redshift can be removed inadequately, and consequently the local environment cannot be accurately measured.
In order to check whether such a situation occurs, we examine whether the size of the redshift slice ($\Delta z$) is larger than the typical error of the photometric redshift ($z_{\rm{err}}$), as follows;
\begin{equation} 
\Delta z  = 0.1 \times (1+z) > z_{\mathrm{err}},
\label{}
\end{equation} 
which is equivalent to the following relation,
\begin{equation} 
\frac{z_{\mathrm{err}}}{1+z} < 0.1.
\label{eq:zerr}
\end{equation} 
Therefore, it is sufficient to check that the photometric redshift and its error satisfy the above equation.
Here $z_{\mathrm{err}}$ is the typical error of the photometric redshift, which is a calculated by the following equation;
\begin{equation} 
z_{\mathrm{err}} = \frac{\mathrm{photoz\_err68\_max - photoz\_err68\_min}}{2},
\end{equation} 
where \texttt{photoz\_err68\_max} and \texttt{photoz\_err68\_min} are the upper and lower limits of the 68\% confidence interval of the photometric redshift.
Figure~\ref{fig:photo-z_err} shows the distribution of $z_{\mathrm{err}}/(1+z)$ of the galaxy sample for the density measurement.
This figure suggests that most of galaxy's photometric redshift error is smaller than the size of the redshift slice, i.e., the photometric errors do not significantly affect the measurement of the local environment.

\begin{figure*}[htbp]
 \centering
  \includegraphics[width=0.7\linewidth]{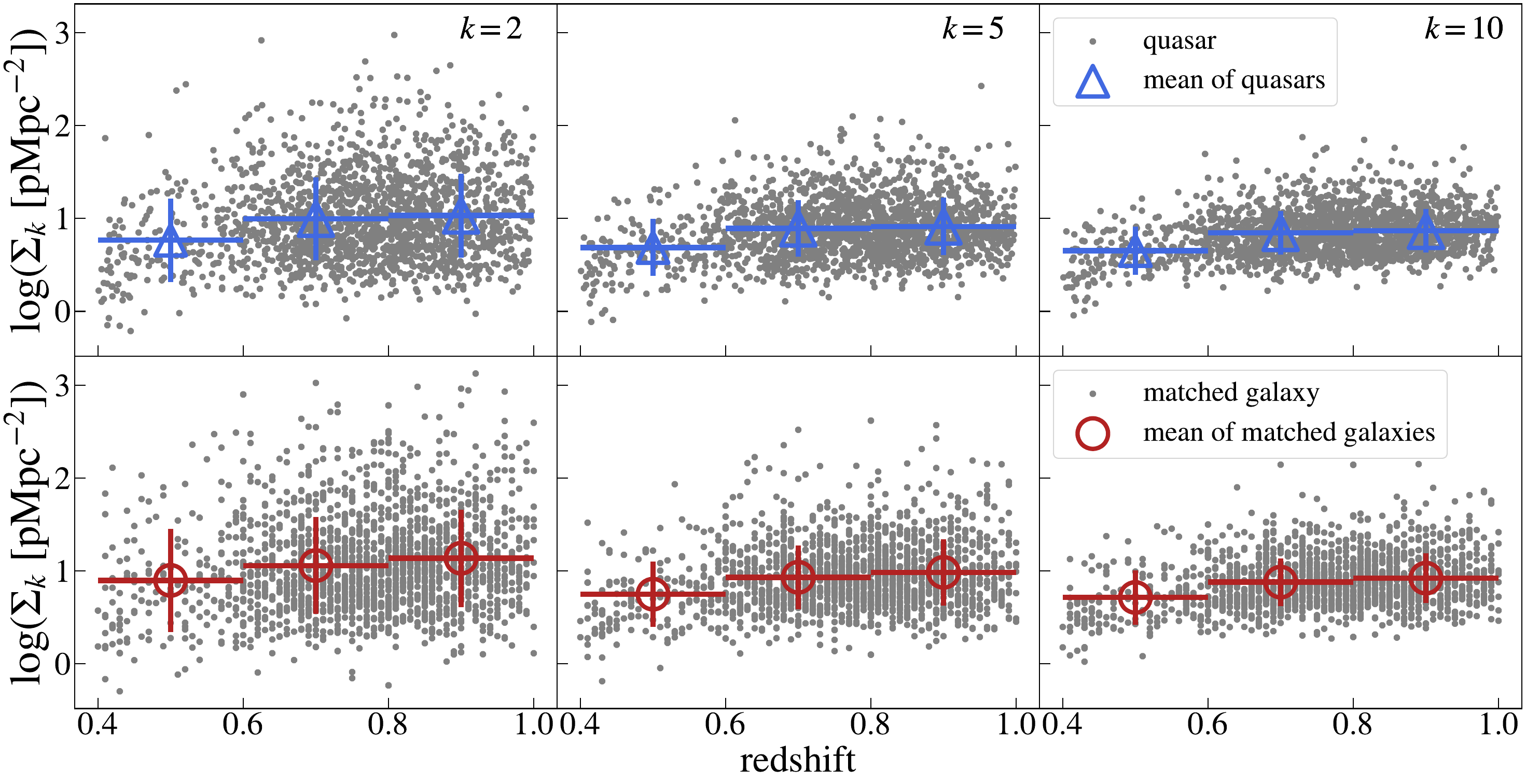}
  \caption{Local number density $ \Sigma_{k} $ with $ k = 2, 5, 10 $ (left, center, and right panels) is shown as a function of the redshift for the quasar sample and the matched galaxy sample. In the top panels, gray dots represent individual quasars, while the blue triangles, vertical bars, and horizontal bars indicate the mean, standard deviation, and range of $ \Sigma_{k} $ for quasars in each redshift bin, respectively. Similarly in the bottom panels, the gray dots represent individual matched galaxies, while the red circles, vertical bars, and horizontal bars indicate the mean, standard deviation, and range of $ \Sigma_{k} $ for matched galaxies in each redshift bin, respectively.}
  \label{fig:density_z}
\end{figure*}

\section{RESULTS} \label{Results}

\subsection{Redshift dependence of the environment}\label{sec:result_density_z}
We show the local galaxy number density $\Sigma_{k}$ for the quasar sample and the matched galaxy sample as a function of redshift in Figure~\ref{fig:density_z}.
The local number density of surrounding galaxies around quasars increases as the redshift increases for all $k$. This relation is also seen for the matched galaxy sample.
The Spearman rank correlation test confirms the statistical significance of the correlation between redshift and $\Sigma_{k}$, showing a significant correlation exceeding 3$\sigma$ for all $k$ in both the quasar and matched galaxy samples (Table~\ref{tbl:qso_matched_dens_spearman_z}).

A plausible main reason for the redshift dependence of \( \Sigma_{k} \) is that the number density of galaxies is intrinsically higher at high redshift (\( z \sim 1.0 \)) than at low redshift (\( z \sim 0.4 \)) due to the evolution of the luminosity function of galaxies. This naturally results in a higher \( \Sigma_{k} \) at a higher redshift, in the redshift range examined in this study.


Here it should be noted that the purpose of this study is to investigate whether there is a systematic difference in the local galaxy environment with and without quasars. To achieve this, as described in Section~\ref{Data}, we construct a one-to-one matched sample of galaxies with similar stellar mass and redshift to the quasar sample. Therefore the redshift dependence of $ \Sigma_{k} $ does not introduce any significant issues in our comparative study about the quasar environment.

\begin{table*}
  \centering
  \caption{Spearman rank correlation test results for local galaxy number density $\Sigma_{k}$ with $k$ = 2, 5, 10 versus the redshift.}
  \label{tbl:qso_matched_dens_spearman_z}
  \begin{threeparttable} 
  \begin{tabular}{lcccccc} 
    \hline
    & \multicolumn{2}{c}{$k=2$} & \multicolumn{2}{c}{$k=5$} & \multicolumn{2}{c}{$k=10$} \\ 
    \hline
                    & $\rho$\tnote{a} & $p$\tnote{b} & $\rho$\tnote{a} & $p$\tnote{b} & $\rho$\tnote{a} & $p$\tnote{b} \\ 
    \hline
    quasar          & $0.14$ & $2.20\times10^{-9}$ & $0.17$ & $6.99\times10^{-13}$ & $0.20$ & $3.31\times10^{-18}$ \\
    matched galaxy  & $0.14$ & $7.25\times10^{-10}$ & $0.17$ & $1.21\times10^{-13}$ & $0.20$ & $7.81\times10^{-18}$ \\ 
    \hline
  \end{tabular}
  \begin{tablenotes}
      \item[a] Correlation coefficient.
      \item[b] $p$-value from the Spearman rank correlation test.
  \end{tablenotes}
  \end{threeparttable}
\end{table*}

\subsection{Comparison of the environment between quasar and matched galaxy samples}\label{sec:result_density}
\begin{figure*}[htbp]
 \begin{center}
  \includegraphics[width=0.9\linewidth]{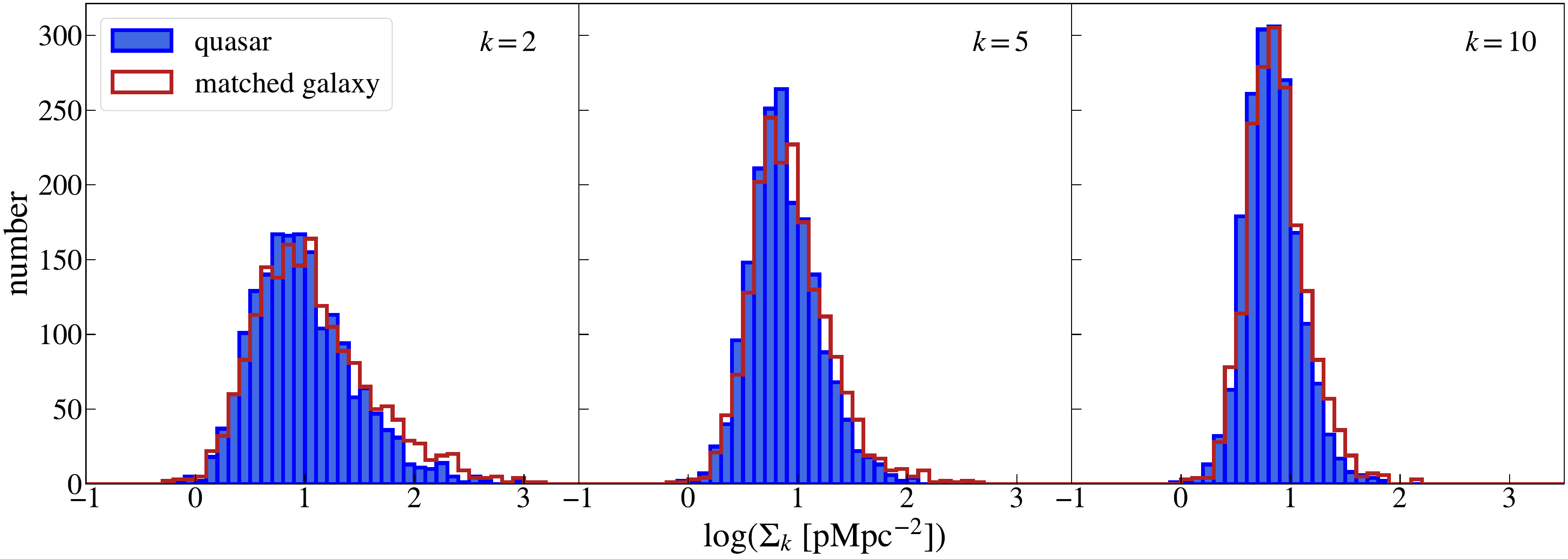}
  \caption{Histograms of local galaxy number densities $\Sigma_{k}$ for quasar sample (blue bar) and matched galaxy sample (thick red line). Left, middle, and right panels show the $\Sigma_{k}$ with $k=2, 5,$ and 10, respectively. }
  \label{fig:density}
 \end{center}
\end{figure*}

We show the derived local galaxy number density $\Sigma_{k}$ for the quasar sample and the matched galaxy sample in Figure~\ref{fig:density}. 
The result suggests that $\Sigma_{k}$ is slightly lower for the quasar sample compared to the matched galaxy sample for all $k\ (=2, 5, 10)$.
We show the statistics of $\Sigma_{k}$ for both samples in Table~\ref{tbl:density}.
The differences in the mean (median) $\Sigma_{k}$ between the quasar sample and the matched galaxy sample are $21.6\%$, $13.2\%$, and $10.9\%$ ($17.8\%$, $10.2\%$, and $8.1\%$) for $k=2, 5$, and 10, respectively.
To evaluate the statistical significance of the differences in the distributions of $\Sigma_{k}$, we perform a KS test for each $\Sigma_{k}$.
The $p$-values for $k=2, 5,$ and $10$ are $1.6\times 10^{-4}$, $2.9\times 10^{-4}$, and $4.2\times 10^{-4}$, respectively. This indicates significant differences in the local environment beyond 3$\sigma$.

\begin{figure*}[htbp]
 \begin{center}
  \includegraphics[width=0.7\linewidth]{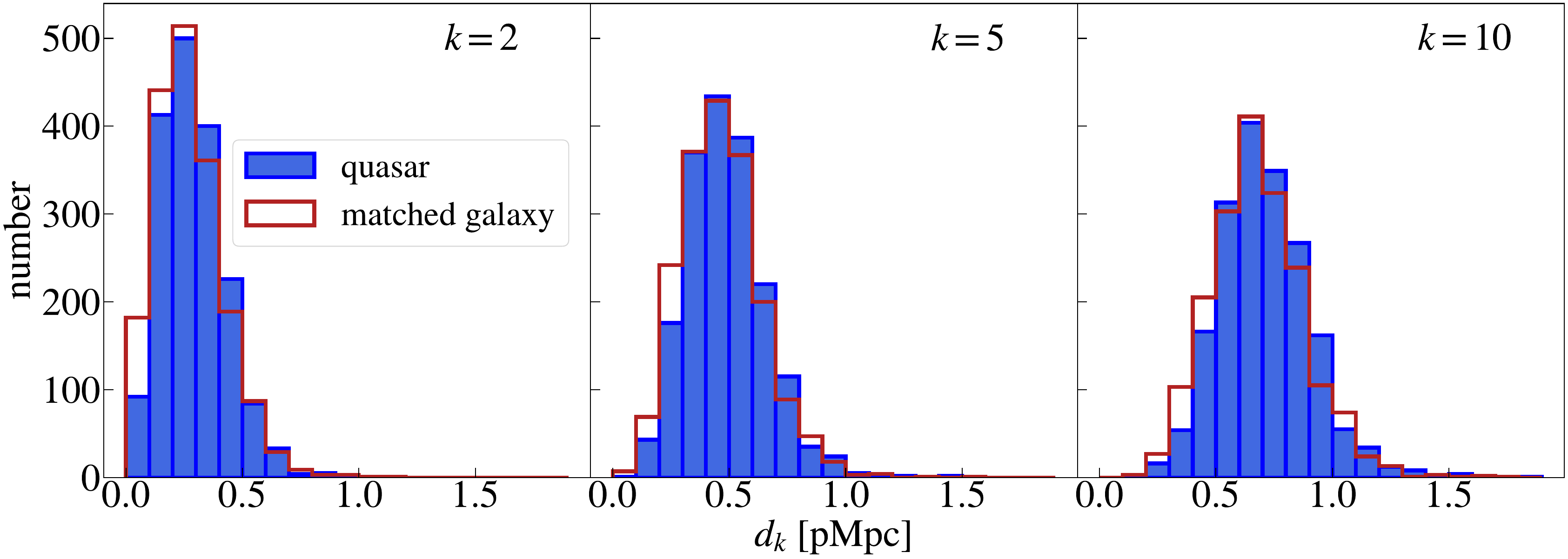}
    \caption{Histograms of the distance to the $k$-th nearest neighbor $d_{k}$ ($k$=2, 5, 10) for quasar sample (blue) and matched galaxy sample (red). Left, central and right panels show the $d_{k}$ with $k=2, 5$, and 10, respectively.}
  \label{fig:qso_dens_qso_dk}
 \end{center}
\end{figure*}

Figure~\ref{fig:qso_dens_qso_dk} shows the histogram of $d_k$ after applying the mask correction. 
This figure demonstrates that our study probes environments on the spatial scale of a few hundreds pkpc.
More specifically, the mean and standard deviation of $d_2$, $d_5$, and $d_{10}$ for the quasar sample are $290 \pm 140~\mathrm{pkpc}$, $490 \pm 170~\mathrm{pkpc}$, and $710 \pm 200~\mathrm{pkpc}$, respectively. For the matched galaxy sample, the mean and standard deviation of $d_2$, $d_5$, and $d_{10}$ are $270 \pm 150~\mathrm{pkpc}$, $460 \pm 170~\mathrm{pkpc}$, and $680 \pm 200~\mathrm{pkpc}$ respectively, suggesting that similar spatial scales are studied for both of the quasar sample and the matched galaxy sample.


\begin{table*}[ht]
  \centering 
  \caption{The mean values, standard deviations, medians, and numbers of local galaxy number densities $\Sigma_{k}$ for quasar sample and matched galaxy sample.}
  \label{tbl:density}
  \hspace*{-2.5cm} 
  \begin{threeparttable}
    \begin{tabular}{lcccccccccccc} 
      \hline \hline
      &\multicolumn{12}{c}{log($\Sigma_{k}$ [pMpc$^{-2}$])}\\ 
      &\multicolumn{4}{c}{$k=2$}	&\multicolumn{4}{c}{$k=5$}	&\multicolumn{4}{c}{$k=10$}\\ \cline{2-13} 
      					& mean & std & median & $N$ & mean & std & median & $N$ & mean & std & median & $N$\\ \hline
      quasar          	& $0.99$ & $0.46$ & $0.93$ & $1760$ & $0.88$ & $0.31$ & $0.85$ & $1817$ & $0.83$ & $0.25$ & $0.82$ & $1847$\\
      matched galaxy  	& $1.07$ & $0.53$ & $1.00$ & $1820$ & $0.93$ & $0.36$ & $0.89$ & $1848$ & $0.88$ & $0.27$ & $0.86$ & $1839$\\\hline
    \end{tabular}
  \end{threeparttable}
\end{table*}

\section{Discussion} \label{Discussion}
\subsection{Comparison with previous studies}
\cite{Karhunen14} studied the environment of 302 quasars in the redshift range $0.1 < z < 0.5$ using data from the SDSS Stripe 82 survey. 
For a comparative study, they constructed a matched sample of inactive galaxies with similar stellar mass and redshift as the quasars and examined the surrounding galaxy environment for both samples. 
\cite{Karhunen14} measured the excess galaxy surface density, or `overdensity', around the targets (both quasars and matched inactive galaxies). 
This overdensity was defined as the ratio of the galaxy density measured near the target to the background galaxy density. 
They reported that the overdensity and uncertainty within 0.1 Mpc of the quasar sample is $1.22 \pm 0.10$, and for the matched inactive galaxy sample is $1.40 \pm 0.11$. 
Similarly, for a larger scale within 0.25~Mpc, the overdensity for quasars is $1.12 \pm 0.04$ while it is $1.21 \pm 0.06$ for the matched sample. \cite{Karhunen14} concluded that there is no statistically significant difference in the environment between the two populations.
However, although not statistically significant, \cite{Karhunen14} reported that the mean number density of surrounding galaxies within 0.1 Mpc and 0.25 Mpc of quasars is lower than that of matched galaxies, by $\sim9.8\%$ and $\sim8.0\%$, respectively.
Our sample size is $\sim$6 times larger than that of \cite{Karhunen14}, which may explain the statistically significant difference confirmed in our study.


\cite{Serber06} conducted an environmental study of 2,028 quasars at $0.08 < z < 0.4$ using SDSS DR3 data. They used a sample of $\sim$10$^5$  $L^*$ galaxies for comparison and reported that the quasar sample resides in significantly denser environment within 1 Mpc than the $L^*$ galaxy sample. 
The overdensity of surrounding galaxies was the highest near quasars ($< 0.1$ Mpc), decreasing gradually out to 1 Mpc. 
Within 0.1 Mpc of the quasars, the overdensity around quasar and uncertainty were reported as $2.12 \pm 0.08$, compared to the $L^*$ galaxies ($1.51 \pm 0.01$), showing a difference of about 30\%. 
Based on these results, \cite{Serber06} concluded that the quasar activity is closely linked to the environment within 0.1 Mpc, suggesting that higher density promotes quasar activity.
Here it should be noted that the comparison between the results by \cite{Serber06} and ours is not straightforward. 
This is because our analysis does not investigate the environment within 0.1 Mpc (see Figure~\ref{fig:qso_dens_qso_dk}), since our target redshift range ($z=0.4$--$1.0$) is higher than that of \cite{Serber06} ($z \sim 0.1$--$0.4$) and thus our analysis is less sensitive to low-luminosity galaxies.
Therefore, the results reported by us and by \cite{Serber06} are not necessarily conflicting.


\cite{Wethers22} conducted an environmental study of 205 quasars at $0.1 < z < 0.35$ using GAMA data, examining the surrounding galaxy environment within 0.1 Mpc. For comparison, they constructed a sample of 20,500 galaxies matched in stellar mass and redshift. They reported that the number of neighbor galaxy within 0.1 Mpc around quasars is $n=0.22\pm0.03$, while the matched galaxy is $n=0.24\pm0.04$. They concluded that the surrounding galaxy environment of the quasar sample is similar to that of the matched galaxy sample. 
In addition to the environment within 0.1 Mpc, \citet{Wethers22} investigated the galaxy environment at a larger scale ($\sim$1 Mpc) using galaxies with an absolute magnitude in the $r$-band of $M_{r} < -20$ and within the redshift range of $z < 0.18$. Due to this redshift limitation, these environment measures were available for only 59 out of the 205 quasars in the \citet{Wethers22} sample. They compared the 5th nearest neighbor for these 59 quasars and 59,000 matched galaxies. Their results showed no significant difference between the quasar and matched galaxy environments, with $\Sigma_{5, \mathrm{QSO}} = 0.75^{+5.32}_{-0.17} \, \mathrm{Mpc^{-2}}$ and $\Sigma_{5, \mathrm{GAL}} = 0.48^{+4.24}_{-0.12} \, \mathrm{Mpc^{-2}}$. 
While their results suggest that quasars tend to reside in higher-density environments, the difference is within the error range. Given the large uncertainties and the limited sample size, further investigation with larger statistical samples is necessary to confirm these findings.

\subsection{Interpretation of the lower-density environment around quasars}
Here, we discuss the physical meanings of our finding that the mean galaxy number density around quasars is lower than that of the matched galaxies.
\cite{Omori23} investigated the environmental dependence of the merger fraction using HSC images of spectroscopic galaxies from the HSC-SDSS and HSC-GAMA crossmatch at $z = 0.01 - 0.35$. They identified a trend in the relationship between galaxy mergers and mass-density environment. Specifically, they found that galaxies with higher merger probability are more commonly found in lower-density environments on scales of 0.5 to 8 $h^{-1}$ Mpc. 
If quasars are mainly triggered by galaxy mergers, their presence in lower-density environments where mergers preferentially occur supports this idea.
This suggests that the lower-density environments around quasars are likely caused by a merger-driven process. Our findings are consistent with this interpretation.
On the other hand, quasars can also be triggered by secular processes.
Such secular processes mostly involve bar or disc instabilities \citep[e.g.,][]{Ohta07, Draper12, Smethurst19}. 
To determine whether quasars are triggered by a merger origin or a secular process origin, it is necessary to conduct a detailed study on even smaller scales than the scale investigate in this study.

\subsection{The environmental dependences of quasar properties}\label{sec:density_property}

\begin{figure*}[htbp]
 \begin{center}
  \includegraphics[width=0.70\linewidth]{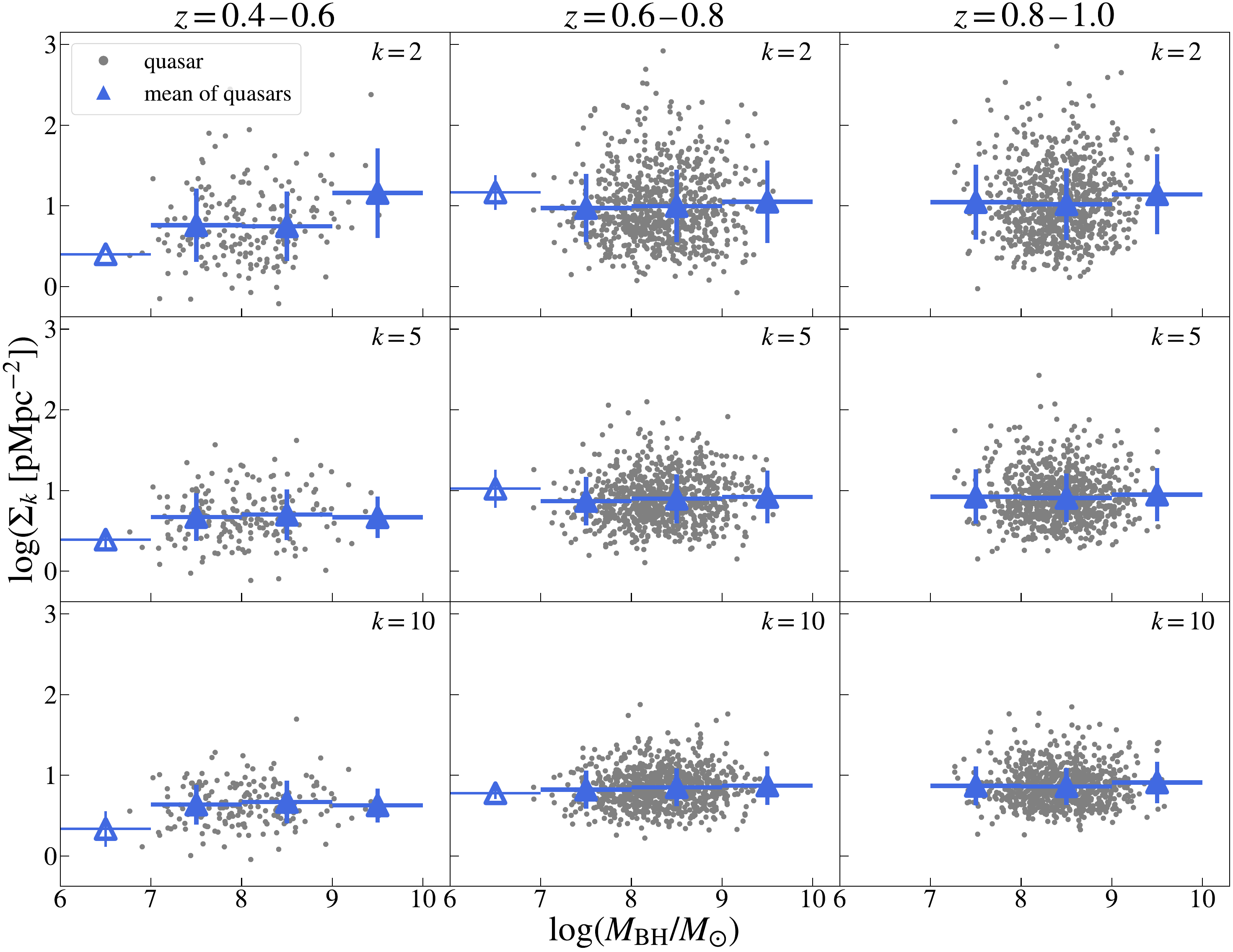}
  \caption{The dependence of the local number densities $\Sigma_{k}$ with $k$=2, 5, 10 (top, middle, and bottom panels) of the quasar sample for each SMBH mass bin. The left, center, and right panels show the data in the redshift ranges $z=0.4$--$0.6$, $0.6$--$0.8$, and $0.8$--$1.0$, respectively. The individual quasars, the mean, and the standard deviation of $\Sigma_{k}$ in each SMBH mass bin are shown in blue dots, blue triangles, and blue bars, respectively. Open triangles indicate the mean for the data with $1 \leq N < 5$. }
  \label{fig:qso_dens_prop}
 \end{center}
\end{figure*}

\begin{figure*}[htbp]
 \begin{center}
  \includegraphics[width=0.70\linewidth]{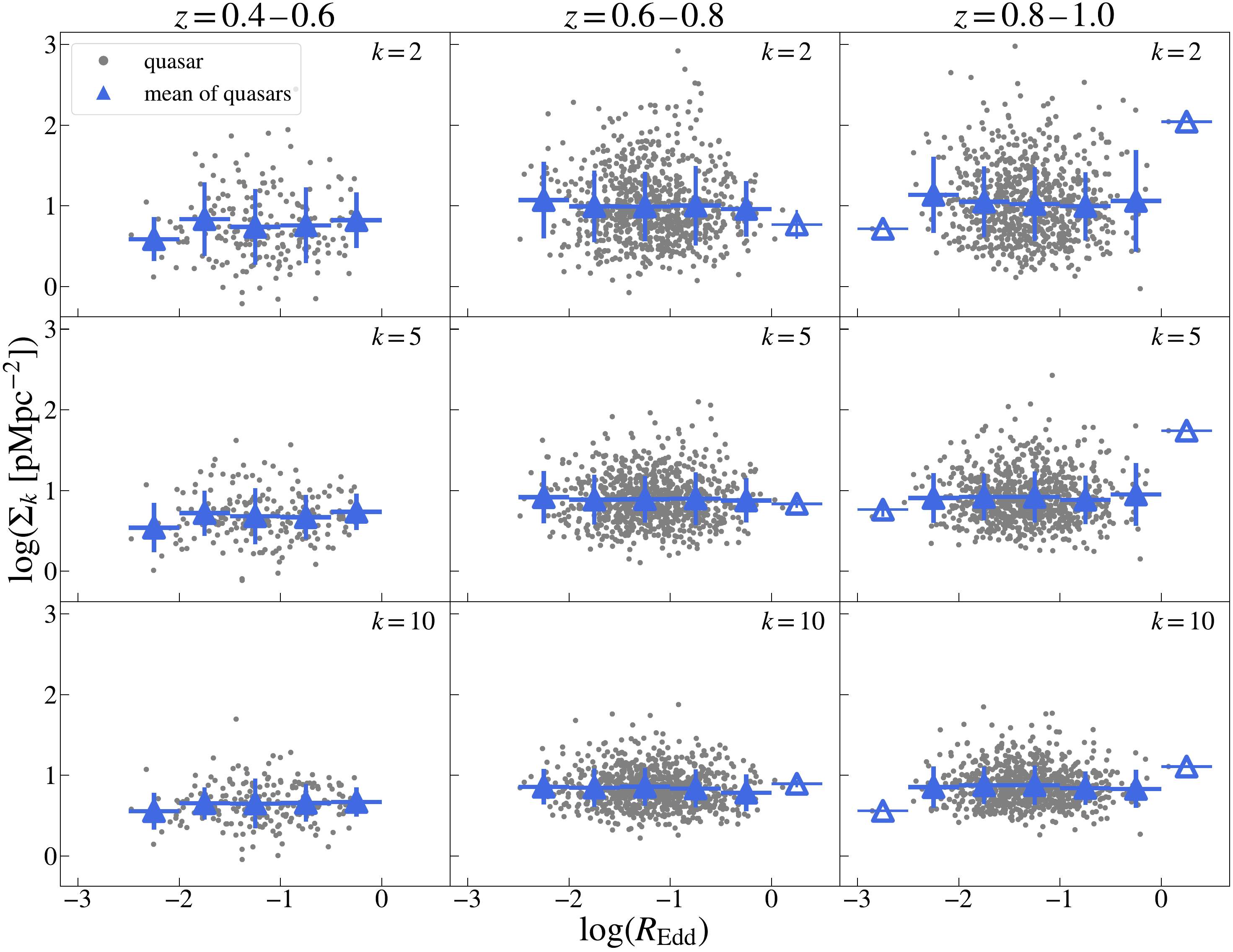}
  \caption{Same and Figure~\ref{fig:qso_dens_prop} but for the dependence on the Eddington ratio. }
  \label{fig:qso_dens_REDD}
 \end{center}
\end{figure*}

As galaxy formation and evolution are heavily influenced by their environment, it is naively expected that some properties of galaxies and their SMBHs are connected to their environment. 
Here, we examine whether the physical properties of quasars are influenced by their surrounding environment.
In Section~\ref{sec:result_density_z}, it is explained that the $\Sigma_{k}$ exhibits a positive correlation with $z$. 
For similar reasons, it can be expected that quasar properties can also show a correlation with $z$. 
In fact, the Spearman rank test for $ z $ vs. $ M_{\mathrm{BH}} $ in our quasar sample shows a correlation coefficient of $\rho = 0.24$ with a $p$-value of $p = 6.9\times10^{-26}$, confirming a statistically significant correlation. 
Similarly, for $ z $ vs. $ R_{\mathrm{Edd}} $, the Spearman rank test shows a correlation coefficient of $ \rho = -0.12 $ with a $p$-value of $ p = 8.9\times10^{-8} $, also indicating a statistically significant correlation.
The relation of $\Sigma_{k}$ and $ M_{\mathrm{BH}} $ for quasars in each redshift bin is shown in Figure~\ref{fig:qso_dens_prop}.
To investigate how $\Sigma_k$ of quasars is related to $M_{\rm BH}$ and $R_{\rm Edd}$ without suffering from the redshift dependence, we divide our quasar sample into the following 3 redshift rages; $ z = 0.4 $--$ 0.6 $, $ 0.6 $--$ 0.8 $, and $ 0.8 $--$ 1.0 $. 
The Spearman rank tests suggests no statistically significant correlation (above 3$\sigma$) between $\Sigma_{k}$ and $ M_{\mathrm{BH}} $ across all redshift bins and scales ($k=2,5,10$) (Table~\ref{tbl:qso_dens_spearman_MBH}).
The statistics of $\Sigma_{k}$ for each of the $ M_{\mathrm{BH}} $ bins are summarized in Table~\ref{tbl:qso_MBH_density}. 
These results indicate that the $M_{\mathrm{BH}}$ is independent of its environment.
We also investigate the relationship between $\Sigma_k$ and $R_{\mathrm{Edd}}$. 
The relationship between $R_{\mathrm{Edd}}$ and the surrounding environment is presented in Figure~\ref{fig:qso_dens_REDD} and Table~\ref{tbl:qso_REDD_density}. 
As shown in Table~\ref{tbl:qso_dens_spearman_REDD}, the Spearman rank test suggests no statistically significant correlation. This means that $R_{\mathrm{Edd}}$ is independent of the local galaxy environment.
Given the independence of $M_{\rm BH}$ and $R_{\rm Edd}$ on the local number density of surrounding galaxies, it is concluded that the properties of quasars do not depend on the local galaxy number density.
This result, combined with the finding in Section~\ref{sec:result_density} that the number density around quasars is lower than that around galaxies, suggests that quasar properties such as $M_{\rm BH}$ and $R_{\rm Edd}$ are more influenced by secular processes rather than the surrounding galaxy environment.


\begin{table}[htbp]
  \centering
  \caption{Results of the Spearman rank correlation test (correlation coffeficient and $p$-value) for local galaxy number density $\Sigma_{k}$ with $k$ = 2, 5, 10 versus the SMBH mass.}
  \label{tbl:qso_dens_spearman_MBH}
  \begin{tabular}{lcccccc} 
    \hline
    & \multicolumn{2}{c}{$k=2$} & \multicolumn{2}{c}{$k=5$} & \multicolumn{2}{c}{$k=10$} \\ 
    \hline
    $z$ range & $\rho$ & $p$ & $\rho$ & $p$ & $\rho$ & $p$ \\ 
    \hline
    $0.4$--$0.6$ & $0.11$ & $0.12$ & $0.07$ & $0.30$ & $0.05$ & $0.45$ \\
    $0.6$--$0.8$ & $0.04$ & $0.22$ & $0.06$ & $0.08$ & $0.06$ & $0.07$ \\
    $0.8$--$1.0$ & $0.04$ & $0.25$ & $0.02$ & $0.59$ & $0.00$ & $0.93$ \\ 
    \hline
  \end{tabular}
\end{table}

\begin{table*}[htbp]
    \centering
    \caption{Local galaxy number density $\Sigma_{k}$ with $k=2, 5, 10$ and numbers of quasars in each SMBH mass bin.}
    \label{tbl:qso_MBH_density}
    \hspace*{-2cm} 
    \begin{tabular}{lcccccccccccc} 
      \hline \hline
       
      &\multicolumn{12}{c}{log($\Sigma_{k}$ [pMpc$^{-2}$])}\\ 
      &\multicolumn{4}{c}{$k=2$}	&\multicolumn{4}{c}{$k=5$}	&\multicolumn{4}{c}{$k=10$}\\ \cline{2-13} 
      					& mean & std & median & $N$ & mean & std & median & $N$ & mean & std & median & $N$\\ \hline
	$z$=0.4--0.6\\ \hline
    $10^{6}  M_{\odot}\text{--}10^{7}  M_{\odot}$ & $0.40$ & $0.02$ & $0.40$ & $2$ & $0.39$ & $0.10$ & $0.39$ & $2$ & $0.34$ & $0.22$ & $0.34$ & $2$\\
	$10^{7}  M_{\odot}\text{--}10^{8}  M_{\odot}$ & $0.76$ & $0.45$ & $0.68$ & $89$ & $0.67$ & $0.3$ & $0.65$ & $93$ & $0.64$ & $0.25$ & $0.63$ & $95$\\
	$10^{8}  M_{\odot}\text{--}10^{9}  M_{\odot}$ & $0.75$ & $0.43$ & $0.68$ & $95$ & $0.70$ & $0.31$ & $0.66$ & $96$ & $0.67$ & $0.26$ & $0.66$ & $96$\\
	$10^{9}  M_{\odot}\text{--}10^{10}  M_{\odot}$ & $1.16$ & $0.56$ & $0.89$ & $7$ & $0.67$ & $0.26$ & $0.6$ & $7$ & $0.63$ & $0.21$ & $0.58$ & $7$\\
       \hline
       $z$=0.6--0.8\\ \hline
       $10^{6}  M_{\odot}\text{--}10^{7}  M_{\odot}$ & $1.17$ & $0.22$ & $1.17$ & $2$ & $1.02$ & $0.24$ & $1.02$ & $2$ & $0.78$ & $0.07$ & $0.78$ & $2$\\
       $10^{7}  M_{\odot}\text{--}10^{8}  M_{\odot}$ & $0.97$ & $0.42$ & $0.90$ & $191$ & $0.87$ & $0.30$ & $0.85$ & $194$ & $0.82$ & $0.24$ & $0.79$ & $192$\\
       $10^{8}  M_{\odot}\text{--}10^{9}  M_{\odot}$ & $1.00$ & $0.45$ & $0.93$ & $534$ & $0.90$ & $0.30$ & $0.86$ & $554$ & $0.85$ & $0.23$ & $0.84$ & $574$\\
       $10^{9}  M_{\odot}\text{--}10^{10}  M_{\odot}$ & $1.05$ & $0.51$ & $1.02$ & $53$ & $0.92$ & $0.33$ & $0.93$ & $54$ & $0.87$ & $0.24$ & $0.85$ & $54$\\
       \hline
       $z$=0.8--1.0\\ \hline
       $10^{6}  M_{\odot}\text{--}10^{7}  M_{\odot}$ & $-$ & $-$ & $-$ & $0$ & $-$ & $-$ & $-$ & $0$ & $-$ & $-$ & $-$ & $0$\\
       $10^{7}  M_{\odot}\text{--}10^{8}  M_{\odot}$ & $1.05$ & $0.46$ & $0.98$ & $125$ & $0.92$ & $0.34$ & $0.85$ & $126$ & $0.87$ & $0.24$ & $0.84$ & $126$\\
       $10^{8}  M_{\odot}\text{--}10^{9}  M_{\odot}$ & $1.02$ & $0.44$ & $0.97$ & $607$ & $0.91$ & $0.30$ & $0.88$ & $628$ & $0.86$ & $0.23$ & $0.84$ & $635$\\
       $10^{9}  M_{\odot}\text{--}10^{10}  M_{\odot}$ & $1.14$ & $0.50$ & $1.08$ & $55$ & $0.95$ & $0.33$ & $0.88$ & $61$ & $0.91$ & $0.26$ & $0.89$ & $64$\\ \hline
     \end{tabular}
\end{table*}

\begin{table*}[htbp]
    \centering
    \caption{Local galaxy number density $\Sigma_{k}$ with $k=2, 5, 10$ and numbers of quasars in each Eddington ratio bin.}
    \label{tbl:qso_REDD_density}
    \hspace*{-2cm} 
    \begin{tabular}{lcccccccccccc} 
      \hline \hline
       
      &\multicolumn{12}{c}{log($\Sigma_{k}$ [pMpc$^{-2}$])}\\ 
      &\multicolumn{4}{c}{$k=2$}	&\multicolumn{4}{c}{$k=5$}	&\multicolumn{4}{c}{$k=10$}\\ \cline{2-13} 
      					& mean & std & median & $N$ & mean & std & median & $N$ & mean & std & median & $N$\\ \hline
	   $z$=0.4--0.6\\ \hline
       $10^{-3.0}\text{--}10^{-2.5}$ & $-$ & $-$ & $-$ & $0$ & $-$ & $-$ & $-$ & $0$ & $-$ & $-$ & $-$ & $0$\\
	   $10^{-2.5}\text{--}10^{-2.0}$ & $0.59$ & $0.27$ & $0.60$ & $10$ & $0.54$ & $0.31$ & $0.54$ & $11$ & $0.56$ & $0.23$ & $0.56$ & $11$\\
	   $10^{-2.0}\text{--}10^{-1.5}$ & $0.83$ & $0.46$ & $0.77$ & $41$& $0.72$ & $0.28$ & $0.66$ & $41$ & $0.65$ & $0.20$ & $0.66$ & $40$\\
	   $10^{-1.5}\text{--}10^{-1.0}$ & $0.74$ & $0.47$ & $0.68$ & $72$ & $0.68$ & $0.35$ & $0.65$ & $72$ & $0.65$ & $0.31$ & $0.64$ & $75$\\
       $10^{-1.0}\text{--}10^{-0.5}$ & $0.76$ & $0.47$ & $0.64$ & $53$ & $0.67$ & $0.28$ & $0.66$ & $56$ & $0.66$ & $0.23$ & $0.63$ & $56$\\
       $10^{-0.5}\text{--}10^{0.0}$ & $0.82$ & $0.35$ & $0.83$ & $17$ & $0.74$ & $0.23$ & $0.69$ & $18$ & $0.67$ & $0.18$ & $0.65$ & $18$\\
       $10^{0.0}\text{--}10^{0.5}$ & $-$ & $-$ & $-$ & $0$ & $-$ & $-$ & $-$ & $0$  & $-$ & $-$ & $-$ & $0$\\
       \hline
       $z$=0.6--0.8\\ \hline
       $10^{-3.0}\text{--}10^{-2.5}$ & $-$ & $-$ & $-$ & $0$ & $-$ & $-$ & $-$ & $0$  & $-$ & $-$ & $-$ & $0$\\
	   $10^{-2.5}\text{--}10^{-2.0}$ & $1.07$ & $0.48$ & $1.18$ & $28$ & $0.92$ & $0.33$ & $0.87$ & $30$ & $0.86$ & $0.22$ & $0.84$ & $31$\\
	   $10^{-2.0}\text{--}10^{-1.5}$ & $0.99$ & $0.44$ & $0.94$ & $148$ & $0.89$ & $0.31$ & $0.85$ & $153$ & $0.85$ & $0.24$ & $0.83$ & $161$\\
	   $10^{-1.5}\text{--}10^{-1.0}$ & $0.99$ & $0.43$ & $0.94$ & $344$ & $0.89$ & $0.29$ & $0.86$ & $357$ & $0.86$ & $0.23$ & $0.84$ & $362$\\
       $10^{-1.0}\text{--}10^{-0.5}$ & $1.00$ & $0.49$ & $0.86$ & $208$ & $0.9$ & $0.33$ & $0.88$ & $213$ & $0.84$ & $0.24$ & $0.82$ & $215$\\
       $10^{-0.5}\text{--}10^{0.0}$ & $0.96$ & $0.35$ & $0.96$ & $50$ & $0.88$ & $0.28$ & $0.85$ & $49$ & $0.78$ & $0.23$ & $0.77$ & $51$\\
       $10^{0.0}\text{--}10^{0.5}$ & $0.77$ & $0.18$ & $0.77$ & $2$ & $0.84$ & $0.05$ & $0.84$ & $2$ & $0.89$ & $0.04$ & $0.89$ & $2$\\
       \hline
       $z$=0.8--1.0\\ \hline
       $10^{-3.0}\text{--}10^{-2.5}$ & $0.72$ & $0.00$ & $0.72$ & $1$ & $0.76$ & $0.09$ & $0.76$ & $2$ & $0.56$ & $0.00$ & $0.56$ & $1$\\
	   $10^{-2.5}\text{--}10^{-2.0}$ & $1.14$ & $0.47$ & $1.07$ & $35$ & $0.91$ & $0.31$ & $0.89$ & $37$ & $0.85$ & $0.25$ & $0.85$ & $40$\\
	   $10^{-2.0}\text{--}10^{-1.5}$ & $1.05$ & $0.44$ & $1.00$ & $243$ & $0.92$ & $0.29$ & $0.89$ & $249$ & $0.88$ & $0.23$ & $0.86$ & $255$\\
	   $10^{-1.5}\text{--}10^{-1.0}$ & $1.02$ & $0.46$ & $0.97$ & $330$ & $0.92$ & $0.32$ & $0.88$ & $341$ & $0.88$ & $0.24$ & $0.85$ & $340$\\
       $10^{-1.0}\text{--}10^{-0.5}$ & $1.00$ & $0.42$ & $0.97$ & $161$ & $0.89$ & $0.30$ & $0.84$ & $169$ & $0.84$ & $0.21$ & $0.83$ & $172$\\
       $10^{-0.5}\text{--}10^{0.0}$ & $1.06$ & $0.63$ & $0.87$ & $16$ & $0.95$ & $0.39$ & $1.00$ & $16$ & $0.83$ & $0.24$ & $0.77$ & $16$\\
       $10^{0.0}\text{--}10^{0.5}$ & $2.04$ & $0.00$ & $2.04$ & $1$ & $1.74$ & $0.00$ & $1.74$ & $1$ & $1.11$ & $0.00$ & $1.11$ & $1$\\ \hline
     \end{tabular}
\end{table*}

\begin{table}[htbp]
  \centering
  \caption{Results of the Spearman rank correlation test (correlation coefficient and $p$-value) for the local galaxy number density $\Sigma_{k}$ with $k$ = 2, 5, 10 versus the Eddington ratio.}
  \label{tbl:qso_dens_spearman_REDD}
  \begin{tabular}{lcccccc} 
    \hline
    & \multicolumn{2}{c}{$k=2$} & \multicolumn{2}{c}{$k=5$} & \multicolumn{2}{c}{$k=10$} \\ 
    \hline
    $z$ range & $\rho$ & $p$ & $\rho$ & $p$ & $\rho$ & $p$ \\ 
    \hline
    $0.4$--$0.6$ & $-0.02$ & $0.80$ & $0.03$ & $0.71$ & $-0.01$ & $0.43$ \\
    $0.6$--$0.8$ & $-0.04$ & $0.26$ & $-0.01$ & $0.78$ & $-0.05$ & $0.16$ \\
    $0.8$--$1.0$ & $-0.04$ & $0.31$ & $-0.02$ & $0.60$ & $0.06$ & $0.43$ \\ 
    \hline
  \end{tabular}
\end{table}

\subsection{Galaxy environment of radio-detected quasars}\label{sec:radio loud qso}
Among quasars, some objects show strong radio emission and are called radio-loud quasars, which are known to account for about 10\% of all quasars \citep[e.g.,][]{Jiang07}. 
It has been also reported that the radio-loud fraction depends on the stellar mass \citep{Best05}.
Previous studies have investigated the relationship between radio-loudness and environment to understand the potential connection between radio emission and environmental factors.
For example, \cite{Falder10} compared the environments of 75 radio-loud quasars and 71 radio-quiet quasars at redshift $ z \sim 1 $ and found that the radio-loud quasars reside in significantly higher-density environments by a factor of $\sim$ 1.6. They argued that the lack of a significant difference in $ M_{\mathrm{BH}} $ between the two samples, coupled with the absence of any difference in their optical color distribution, suggests no direct relationship between $ M_{\mathrm{BH}} $ and radio strength. 
Furthermore, they found that the number density of surrounding galaxies increases with the radio luminosity of AGN. This implies that the radio emission of quasars is influenced by their surrounding galaxy environment.
On the other hand, \cite{Coziol17} investigated the environments of 431 radio-detected and 1,527 radio-undetected quasars, and they reported no significant differences in their environment.
Studies by \cite{Karhunen14} and \cite{Wethers22} also found no significant differences in the galaxy environments of quasars with and without radio detections.
Thus, the relationship between the radio emission of quasars and their environment remains controversial.
Therefore, we divide our quasar sample into radio-detected and radio-undetected groups and examine their surrounding galaxy environment to determine whether the radio emission of quasars is related to the environment.

The quasar catalog used in this study \citep{Paris18, Rakshit20} includes information of the radio detection. 
Specifically, a quasar is classified as a radio-detection object when radio source from the Faint Images of the Radio Sky at Twenty cm (FIRST) catalog \citep{Becker95,Helfand15} is found within 2$^{\prime\prime}$ of the SDSS coordinate.
Among the 1,912 quasars, 50 are classified as radio-detected quasars (6 at $z = 0.4$--$0.6$, 12 at $z = 0.6$--$0.8$, and 32 at $z = 0.8$--$1.0$).
The mean and standard deviation of the redshift, BH mass, $i$-band absolute magnitude, and Eddington ratio of the radio-detected 50 quasars are $z = 0.80\pm0.15$, $\mathrm{log(}M_{\mathrm{BH}}/M_{\odot}) = 8.72\pm0.44$, $M_{i} = -23.85\pm1.04$, and $R_{\mathrm{Edd}} = -1.43\pm0.46$, respectively.
The mean and standard deviation of the redshift, SMBH mass, $i$-band absolute magnitude, and Eddington ratio of the radio-undetected 1862 quasars are $z = 0.77\pm0.13$, $\mathrm{log(}M_{\mathrm{BH}}/M_{\odot}) = 8.32\pm0.45$, $M_{i} = -23.28\pm0.76$, and $R_{\mathrm{Edd}} = -1.25\pm0.44$, respectively.
The mean and standard deviation of local galaxy densities, as well as the number of objects meeting the masking criteria for both radio-detected and radio-undetected quasars, are given in Table~\ref{tbl:density_radio}. 
As shown in Table~\ref{tbl:density_radio}, the differences in local galaxy density between radio-detected and radio-undetected quasars are 13\%, 20\%, and 11\% for $k=2, 5$, and 10, respectively.
To examine the statistical significance of possible differences in the local galaxy density $ \Sigma_{k} $ (for $ k = 2, 5, 10 $) between radio-detected and radio-undetected quasars, we perform the KS test. The results show $p$-values of 0.76, 0.36, and 0.50 for $k=2, 5$, and 10, respectively, indicating no statistically significant difference in the environments based on the radio detection.

As mentioned above, previous studies have shown that the radio-loud fraction depends on the stellar mass (or SMBH mass).
Our analysis finds no significant correlation between the SMBH mass and the local galaxy number density (see Section~\ref{sec:density_property}).
These two facts naturally explain the absence of a significant environmental difference between radio-detected and radio-undetected quasars.
Our findings are also consistent with those of \citet{Coziol17}, \citet{Karhunen14}, and \citet{Wethers22}, but inconsistent with the results of \citet{Falder10}. \citet{Coziol17} compared the environments of radio-detected and radio-undetected quasars, whereas \citet{Falder10} investigated the environments of radio-loud and radio-quiet quasars. Since our analysis also focuses on the comparison between radio-detected and radio-undetected quasars, the difference in classification may explain the discrepancy. 
Another possible reason is the small sample size of radio-detected quasars in our study, which results in less robust statistical conclusions.


\begin{table*}[t]
  \centering 
  \caption{The local number densities $\Sigma_{k}$ with $k$=2, 5, 10 and numbers of radio-detected quasar sample and radio-undetected quasar sample.}
  \label{tbl:density_radio}
  \hspace*{-2.5cm} 
    \begin{tabular}{lcccccccccccc} 
      \hline \hline
      &\multicolumn{12}{c}{log($\Sigma_{k}$ [pMpc$^{-2}$])} \\ 
      &\multicolumn{4}{c}{$k=2$}	&\multicolumn{4}{c}{$k=5$}	&\multicolumn{4}{c}{$k=10$} \\ \cline{2-13} 
      							& mean & std & median & $N$ & mean & std & median & $N$ & mean & std & median & $N$ \\ \hline
      radio-detected quasar    	& $1.04$ & $0.49$ & $0.95$ & $45$  & $0.96$ & $0.32$ & $0.91$ & $46$  & $0.88$ & $0.23$ & $0.87$ & $49$ \\
      radio-undetected quasar  	& $0.99$ & $0.46$ & $0.93$ & $1715$ & $0.88$ & $0.31$ & $0.85$ & $1771$ & $0.83$ & $0.25$ & $0.82$ & $1798$ \\ \hline
    \end{tabular}
\end{table*}

\section{Conclusion}\label{Conclusion}
We investigate the environments of SDSS DR14 quasars at $z=0.4$--$1.0$ to understand whether quasars live in high-density or low-density regions compared to matched galaxies, using the wide and deep observational data of HSC-SSP. 
We adopt the $k$-nearest neighbor method to define the local galaxy environment around quasars and matched galaxies and compare their respective environments. 
Furthermore, within the quasar sample, we investigate possible correlations between the environment and quasar properties such as SMBH mass and Eddington ratio.
In addition, we also compare the environment of radio-detected and radio-undetected quasars.

Our findings in this study are as follows:
\begin{enumerate}
    \item The  local number density of galaxies around quasars is significantly correlated with the redshift. The main reason of this relationship is the intrinsic increase in the number density of galaxies with brighter absolute magnitudes at high redshift compared to low redshift.
    \item The mean local number density for quasars is lower than that of matched galaxies by $\sim11$--$20\%$ in the spatial scale of 0.3--0.7 pMpc.
    \item The correlations are not found between the local galaxy number density and quasar properties such as the SMBH mass and Eddington ratio, by separating the quasars into different redshift bins.
     \item There is no significant difference in the environment between radio-detected and radio-undetected quasars.
\end{enumerate}

In the future, the large-scale spectroscopic survey conducted by Subaru Prime Focus Spectrograph \citep{Takada14, Greene22} will enable spectroscopic observation of faint galaxies around quasars, allowing for more detailed studies of galaxy environments at closer scales around quasars. 
Additionally, the DESI \citep{2016arXiv161100036D} and MOONS \citep{Cirasuolo11, Cirasuolo14} surveys will also provide spectroscopic data, helping to gather more detailed redshift information and further enhancing our understanding of galaxy environments.
These data will enable a more detailed investigation of the galaxy environment.

\section*{acknowledgments}
This work is supported by Japan Science and Technology Agency (JST) Support for Pioneering Research Initiated by the Next Generation (SPRING), Grant Number JPMJSP2162. 
This work is financially supported by the Japan Society for the Promotion of Science (JSPS) KAKENHI grants (20H01949, 21H04496, and 23H01215; T.N., 23K22537; Y.T.). 
\vspace{0.5\baselineskip}

The HSC collaboration includes the astronomical communities of Japan and Taiwan, and Princeton University.  The HSC instrumentation and software were developed by the National Astronomical Observatory of Japan (NAOJ), the Kavli Institute for the Physics and Mathematics of the Universe (Kavli IPMU), the University of Tokyo, the High Energy Accelerator Research Organization (KEK), the Academia Sinica Institute for Astronomy and Astrophysics in Taiwan (ASIAA), and Princeton University.  Funding was contributed by the FIRST program from the Japanese Cabinet Office, the Ministry of Education, Culture, Sports, Science and Technology (MEXT), the Japan Society for the Promotion of Science (JSPS), Japan Science and Technology Agency  (JST), the Toray Science  Foundation, NAOJ, Kavli IPMU, KEK, ASIAA, and Princeton University.
This paper is based on data collected at the Subaru Telescope and retrieved from the HSC data archive system, which is operated by Subaru Telescope and Astronomy Data Center (ADC) at NAOJ. Data analysis was in part carried out with the cooperation of Center for Computational Astrophysics (CfCA) at NAOJ.  We are honored and grateful for the opportunity of observing the Universe from Maunakea, which has the cultural, historical and natural significance in Hawaii.
 This paper makes use of software developed for Vera C. Rubin Observatory. We thank the Rubin Observatory for making their code available as free software at http://pipelines.lsst.io/. 
 The Pan-STARRS1 Surveys (PS1) and the PS1 public science archive have been made possible through contributions by the Institute for Astronomy, the University of Hawaii, the Pan-STARRS Project Office, the Max Planck Society and its participating institutes, the Max Planck Institute for Astronomy, Heidelberg, and the Max Planck Institute for Extraterrestrial Physics, Garching, The Johns Hopkins University, Durham University, the University of Edinburgh, the Queen's University Belfast, the Harvard-Smithsonian Center for Astrophysics, the Las Cumbres Observatory Global Telescope Network Incorporated, the National Central University of Taiwan, the Space Telescope Science Institute, the National Aeronautics and Space Administration under grant No. NNX08AR22G issued through the Planetary Science Division of the NASA Science Mission Directorate, the National Science Foundation grant No. AST-1238877, the University of Maryland, Eotvos Lorand University (ELTE), the Los Alamos National Laboratory, and the Gordon and Betty Moore Foundation.

\appendix 
We use some flags to select a clean samples. Below, we provide details on the HSC galaxy selection flags (Appendix \ref{Ap:A}) and SDSS quasar selection flags (Appendix \ref{Ap:B}).

\section{The HSC Galaxy Selection Flags}\label{Ap:A}

\begin{quote}
  \begin{itemize}
    \item $0.20 < \mathrm{photoz\_best} < 1.20$\\
      Redshift range for the galaxy sample.
    \item $\mathrm{photoz\_risk\_best} \leq 0.1$\\
      No concern about the risk of $ \mathrm{photoz\_best} $ falling outside the range $ z_\mathrm{best} \pm 0.15(1+z_\mathrm{best}) $.
    \item $\mathrm{reduced\_chisq} \leq 3.0$\\
    	Reduced chi-squares of the best-fit template.
    \item $\mathrm{i\_cmodel\_mag} - a\_i \leq 23.5$ \\ 
    	To exclude high-redshift galaxies, we select galaxies which brighter than 23.5 mag.
    \item $(g, r, i, z, y)\_\mathrm{inputcount\_value} \geq$ (3, 3, 5, 5, 5) \\
    	Number of images contributing.
	Number of images contributing is 3 or more in each of the $g$-band and $r$-band and 5 or more in each of the $i$-band, $z$-band, and $y$-band.
    \item $\mathrm{isprimary} = \mathrm{True}$\\
    	A source has no children.
    \item $(g, r, i, z, y)\_\mathrm{cmodel\_flag} = \mathrm{False}$\\
    	Cmodel measurement has no problem.
    \item $(g, r, i, z, y)\_\mathrm{pixelflags\_edge} = \mathrm{False}$\\
    	Located within images.
    \item $(g, r, i, z, y)\_\mathrm{pixelflags\_interpolatedcenter} = \mathrm{False}$\\  
    	None of the central pixels of an object is interpolated.
    \item $(g, r, i, z, y)\_\mathrm{pixelflags\_saturatedcenter} = \mathrm{False}$\\
    	None of the central pixels of an object is saturated.
    \item $(g, r, i, z, y)\_\mathrm{pixelflags\_crcenter} = \mathrm{False}$\\  
    	None of the central pixels of an object is affected by a cosmic ray.
    \item $(g, r, i, z, y)\_\mathrm{pixelflags\_bad} = \mathrm{False}$\\
    	None of the pixels in the footprint of an object is labelled as bad.
    \item $(g, r, i, z, y)\_\mathrm{mask\_brightstar\_halo} = \mathrm{False}$\\  
    	The source is not within the halo mask of a bright star.
    \item $(g, r, i, z, y)\_\mathrm{mask\_brightstar\_ghost} = \mathrm{False}$\\  
    	The source is not within the ghost mask of bright star.
    \item $(g, r, i, z, y)\_\mathrm{mask\_brightstar\_blooming} = \mathrm{False}$\\  
    	The source is not within the blooming mask of bright star.
    \item $(y)\_\mathrm{mask\_brightstar\_scratch} = \mathrm{False}$\\  
    	The source is not within the scratch mask of bright star.
    
  \end{itemize}
\end{quote}

\section{The SDSS Quasar Selection Flags}\label{Ap:B}
\begin{quote}
 \begin{itemize}
  \item $0.40 \leq \rm{Z\_VI} \leq 1.00$ \\
    Redshift range for the quasar sample.
  \item $-30.0 < M_i < -10.0$ \\
    Absolute $i$-band magnitude limits.
  \item $\rm{ZWARNING} == 0$ \\
    Classification and redshift are considered reliable.
  \item $\rm{LOG \_ MBH} \neq -999$ \\
    Excludes objects where the SMBH mass is not measured or unreliable.
  \item $\rm{LOG \_ LOBL} \neq -999$ \\
    Filters out quasars with missing or unreliable broad-line region luminosities.
  \item $\rm{LOG \_ MBH\_ERROR} < 0.5$ \\
    Restricts the sample to quasars with reasonable SMBH mass measurements.
  \item $\rm{(Z\_VI} < 0.80)\ \& \ \rm{(FWHM \_ HB \_ BR} < 15000)$ \\
    For quasars with $z < 0.8$, a limit is imposed on the broad-line width of H$\beta$ to improve the accuracy of SMBH mass measurements.
  \item $\rm{(Z\_VI} \geq 0.80)\ \& \ \rm{(FWHM \_ MGII \_ BR} < 15000)$ \\
    For quasars with $ z \geq 0.8 $, a limit is imposed on the broad-line width of MgII to improve the accuracy of SMBH mass measurements.
  \item $\rm{ANCILLARY \_ TARGET2} \neq 25 \rm{\ or\ } 27 \rm{\ or\ } 32 \rm{\ or\ } 56$ \\
    Excludes quasars flagged as belonging to certain ancillary target categories to maintain sample homogeneity. These include target 25 (SPIDERS\_PILOT), target 27 (QSO\_VAR\_LF), target 32 (XMM\_PRIME), and target 56 (QSO\_DEEP). For more details, see Table 1 of Paris et al. (2018). \label{eq:ANC}
 \end{itemize}
\end{quote}
Note: The flag of ANCILLARY\_TARGET2 is for the selection criteria of SDSS quasar spectroscopic targets. Without applying this condition, regions with partially higher quasar number density may arise.


\bibliography{Ref}{}
\bibliographystyle{aasjournalv7}



\end{document}